\documentclass[]{article}
\usepackage[margin=3cm]{geometry}
\usepackage{color}
\usepackage{url}
\usepackage{graphicx}
\usepackage{geometry}
\usepackage{lipsum}
\usepackage[export]{adjustbox}
\usepackage[font=footnotesize,labelfont=bf]{caption}
\usepackage{mathtools} 
\usepackage[T1]{fontenc}
\usepackage{gensymb}
\usepackage{amsmath}
\usepackage{amssymb}
\usepackage{fancyhdr}
\usepackage[utf8]{inputenc}
\usepackage{authblk}
\usepackage{rotating}
\usepackage{comment}
\usepackage[dvipsnames]{xcolor}
\usepackage{authblk}
\usepackage{cite}
\usepackage{url}

\title{Steady-state fluctuations of a genetic feedback loop \\ with fluctuating rate parameters using the \\ unified colored noise approximation}
\author[1]{James Holehouse$^{* 1}$, Abhishek Gupta\thanks{These authors contributed equally.}$^{1}$ and Ramon Grima\thanks{Correspondence: Ramon.Grima@ed.ac.uk.}}
\affil[1]{School of Biological Sciences, University of Edinburgh, U.K.}
\date{\today}

\begin{document}

\maketitle

\begin{abstract}
    A common model of stochastic auto-regulatory gene expression describes promoter switching via cooperative protein binding, effective protein production in the active state and dilution of proteins. Here we consider an extension of this model whereby colored noise with a short correlation time is added to the reaction rate parameters -- we show that when the size and timescale of the noise is appropriately chosen it accounts for fast reactions that are not explicitly modelled, e.g., in models with no mRNA description, fluctuations in the protein production rate can account for rapid multiple stages of nuclear mRNA processing which precede translation in eukaryotes. We show how the unified colored noise approximation can be used to derive expressions for the protein number distribution that is in good agreement with stochastic simulations. We find that even when the noise in the rate parameters is small, the protein distributions predicted by our model can be significantly different than models assuming constant reaction rates. 
\end{abstract}


\section{Introduction}
Proteins perform a large range of cellular functions and hence it is of great interest to understand how the genes that produce them operate. Autoregulation is a  mechanism to regulate gene expression whereby proteins expressed by a certain gene can subsequently bind to the same gene and cause an increase or a decrease in its expression (positive and negative feedback, respectively) \cite{alberts2002molecular}. Autoregulation is common; for example in \textit{E. coli} it is estimated that 40\% of all transcription factors are self-regulated \cite{shen2002network,rosenfeld2002negative}.

For almost two decades, it has been known that gene expression is inherently stochastic \cite{elowitz2002stochastic}, and as such the modelling of gene regulatory networks must account for this stochasticity. Noise in gene expression can be split into two contributions: (i) Fluctuations in the number of messenger RNA (mRNA) and proteins due to inherent randomness in the time at which the processes of transcription and translation occur. This is often described as intrinsic noise. For example, if a gene is always active and produces mRNA with rate parameter $k$ then the time between successive transcription events is an exponentially distributed random variable with mean $1/k$. (ii) Fluctuations in the number of mRNA and proteins due to fluctuations in the rate parameters themselves. For example the transcription rate $k$ mentioned previously, while often taken to be a constant, it is generally a fluctuating quantity itself, because of fluctuating numbers of polymerases and transcription factors. Another example would be fluctuations in the protein numbers due to fluctuations in the translation rate stemming from fluctuating numbers of ribosomes in the cell \cite{elowitz2002stochastic,swain2002intrinsic}. This noise is often termed extrinsic noise.

The division of noise into these two categories is of course artificial but it is useful from a conceptual and modelling point of view. The simulation of stochastic biochemical processes is most commonly done using the stochastic simulation algorithm (SSA) \cite{gillespie1977exact} which assumes that the rate parameter of a reaction will not change in the interval between two successive reaction events. Hence the prevalent means of stochastic simulation assumes that noise is either principally intrinsic or that if there is extrinsic noise it is operating on long timescales such that it is slowly varying. While this may be the case sometimes, it is not generally true. This is because whenever we have an effective  reaction that lumps together a large number of intermediate reactions (a multi-stage reaction process), we are making the inherent assumption that these intermediate reactions occur very fast and hence naturally the effective rate parameter is fluctuating on a fast timescale. 

Taking into account these fluctuations is however not a simple feat. The chemical master equation (CME, \cite{gardiner1985handbook,van1992stochastic}) describing the Markov process simulated by the SSA has been solved exactly or approximately to obtain the protein number distribution in steady-state for a wide variety of models of autoregulation  \cite{holehouse2020stochastic,grima2012steady,kumar2014exact,liu2016decomposition,jia2019single,jia2020small,cao2018linear,kurasov2018stochastic,andreychenko2017distribution,thomas2012slow,friedman2006linking,ochab2015transcriptional,jkedrak2016influence}, provided the rate parameters are assumed to be constant, i.e., ignoring extrinsic noise. There are however a number of studies that have analyzed stochastic models with fluctuating rate parameters, which we summarize next. Modifications of the linear noise approximation (a type of Fokker-Planck approximation of the CME) incorporating extrinsic noise have proved popular to approximate moments for systems subject to small magnitudes of extrinsic noise with certain properties: (i) for time-independent Gaussian colored noise \cite{scott2006estimations,toni2013combined} and (ii) more realistic lognormally distributed noise \cite{keizer2019extending}. Wentzel-Kramers-Brillouin (WKB) methods have also been utilised for cases where the correlation time of the colored noise is tending either to zero or to infinity \cite{roberts2015dynamics}. These methods provide probability distributions for systems where the noise on the rate parameters is drawn from a negative binomial distribution, however their analysis does not easily translate to finding good approximations for steady states probability distributions where the correlation time of colored noise is neither small or large.

The focus of the present article is threefold: (i) to provide a general method by which one can obtain analytical expressions for the steady-state protein distributions of auto-regulatory gene circuits with fluctuating rate parameters, through the use of the \textit{unified colored noise approximation} (UCNA) \cite{jung1987dynamical}, (ii) to use this method to investigate the effects that extrinsic noise of different magnitude and timescales has on auto-regulatory gene expression and (iii) to show how the colored noise formalism can be used to describe complex models of autoregulation that involve multi-stage protein production and multi-stage protein degradation. We note that the UCNA was previously utilised in a gene expression context \cite{shahrezaei2008colored} for linear reaction networks that are deterministically monostable and in which there is no feedback mechanism. Our analysis goes further, exploring the addition of colored noise to a non-linear reaction network which expresses deterministic bistability, whilst also incorporating intrinsic fluctuations from the core gene expression processes.

The structure of our paper is as follows. In Section \ref{sec2} we introduce the cooperative auto-regulatory reaction scheme that we will study in this article. We also show that for non-fluctuating rate parameters, the analytical protein distribution given by the chemical Fokker-Planck equation provides an excellent approximation of the protein distribution solution of the CME, in the limit of fast gene switching. In Section \ref{sec3} we add colored noise to each reaction in the auto-regulatory reaction scheme (assuming fast gene switching) and use the UCNA to derive the protein number distribution solution of the chemical Fokker-Planck equation. The solution is shown to be in good agreement with a stochastic simulation algorithm modified to account for extrinsic noise on the rate parameters. We also use the solution to investigate the effect that extrinsic noise has on the number of modes of the protein distribution and clarify the limits of the UCNA derivation, including the three main conditions which cause it to breakdown. In Section \ref{cUCNA} we extend the analysis to the limit of slow gene switching by introducing a conditional version of the UCNA. In Section \ref{sec:app} we show two examples of how one can successfully model complex auto-regulatory systems by means of simpler ones with colored noise on the reaction rate parameters, here done for multi-stage protein production and multi-stage degradation. We conclude in Section \ref{conc} with a discussion of our results and further problems to be addressed on this topic.

\section{Approximate solution for autoregulation with non-fluctuating rate parameters}\label{sec2}
We consider the reaction scheme for a genetic non-bursty cooperative feedback loop, where for simplicity we neglect the presence of mRNA:
\begin{align}\label{eq:coopRS}
    G \xrightarrow[]{r_u}G + P,\, G^* \xrightarrow[]{r_b} G^* + P,\, G+2P \xrightleftharpoons[s_u]{s_b} G^*,\, P \xrightarrow{d} \varnothing.
\end{align}
The reactions $G \xrightarrow[]{r_u}G + P$ and $G^* \xrightarrow[]{r_b} G^* + P$ model the production of protein in each gene state, $G+2P \xrightleftharpoons[s_u]{s_b} G^*$ models the binding and unbinding of the gene to the proteins (with cooperativity 2), and $P \xrightarrow{d} \varnothing$ models the dilution/degradation of proteins inside the cell. It is assumed there is only one gene present in the system and hence we are either in state $G$ or $G^*$ at any one time. Before considering the addition of colored noise to the reaction rate parameters above, we first consider the solution with constant rate parameters to provide a reference point for approximations made in Section \ref{sec3}, and to clarify the approximation of a CME by a one variable chemical Fokker-Planck equation (FPE).

The CME for the reaction scheme in Eq. (\ref{eq:coopRS}) does not have a known exact solution, even at steady-state for constant reaction rate parameters, and so approximations are necessary. Note that in what follows, we will use the terminology ``reaction rate parameters'' and ``rates'' interchangeably. We first consider the limit of fast gene switching -- i.e., the frequency of gene activation and inactivation events is much larger than the frequency of any other reaction in the system. Later in Section \ref{cUCNA} we will discuss approximations for the slow switching limit. Where $[g^*]$ and $[g]$ are the deterministic mean number of bound and unbound gene respectively and $[n]$ is the mean protein number, the rate equations for the reaction scheme in Eq. (\ref{eq:coopRS}) are:
\begin{align}\label{re1}
    \frac{d [g]}{dt} &= s_u [g^*] - \frac{s_b}{\Omega^2} [g] [n]^2,\\\label{re2}
    \frac{d [n]}{dt} &= 2 s_u [g^*] - \frac{2 s_b}{\Omega^2} [g] [n]^2 + r_u[g]+r_b[g^*]-d[n],
\end{align}
where $[g]+[g^*]=1$. For clarity we state the units of each rate parameter: $r_u$, $r_b$, $d$ and $s_u$ have units of $\text{s}^{-1}$, and $s_b$ has units of $\text{Volume}^{2}\cdot\text{s}^{-1}$. This ensures a matching of the units with the left hand side of Eqs. (\ref{re1}--\ref{re2}), which has units of $\text{s}^{-1}$. In the fast switching limit, the gene rapidly equilibrates to quasi-steady state conditions, i.e. $d[g]/dt \approx d[g^*]/dt \approx 0$ and hence the deterministic rate equation for mean protein number reduces to a much simpler form:
\begin{align}\label{eq:effre}
    \frac{d [n]}{d t} = \frac{L r_u + r_b ([n]/\Omega)^2}{L+([n]/\Omega)^2}-d [n],
\end{align}
where $L = s_u/s_b$. Note that the reaction scheme here described exhibits deterministic bistability over some regions of the parameter space. This equation is consistent with a birth-death process where proteins are produced via a zeroth-order reaction (which is dependent on the number of proteins) and are destroyed by a first-order reaction \cite{holehouse2019revisiting}. The CME for this reduced process is given by:
\begin{align}\label{eq:redMEnb}
    \frac{d P_a(n,t)}{dt} = T^{+}(n-1)P_a(n-1,t) + T^{-}(n+1)P_a(n+1,t) - (T^{+}(n)+T^{-}(n))P_a(n,t),
\end{align}
where $P_a(n,t)$ is the probability that at a time $t$ there are $n$ proteins in the system; $T^{+}(n)$ and $T^{-}(n)$ are the propensities of protein production and degradation respectively. The subscript $a$ denotes that this is the probability for the reduced system, an \textit{approximate} solution to the master equation of the full system. $T^{+}(n) dt$ is the probability, given $n$ proteins are in the system, that a protein production reaction occurs, increasing the protein number of the system by 1, in the time interval $[t,t+dt)$. Similarly, $T^{-}(n) dt$ is the probability, given $n$ proteins are in the system, that a protein degradation event occurs, decreasing the protein number by 1, in the time interval $[t,t+dt)$. These propensities are given by:
\begin{align}\label{eq:Propsnb}
    T^{+}(n) &= \frac{r_u L + r_b (n/\Omega)^2}{L+(n/\Omega)^2},\\\label{eq:Propsnb2}
    T^{-}(n) &= d \,n.
\end{align}
These propensities are deduced directly from the form of the effective rate equation in Eq. (\ref{eq:effre}). Essentially, the probability for a particular reaction per unit time is taken to be the same as the reaction rate in the effective deterministic rate equation with $[n]$ replaced by $n$. {\textit{We emphasise that while this appears to be a heuristic rule with no apparent fundamental microscopic basis, it has been shown that the reduced master equation based on it provides an accurate approximation to the SSA of the full reaction system in fast gene switching conditions provided the low protein number states are rarely visited \cite{jia2020small,holehouse2019revisiting}}}.

\begin{figure}[h!]
\centering
\includegraphics[width=0.5\textwidth]{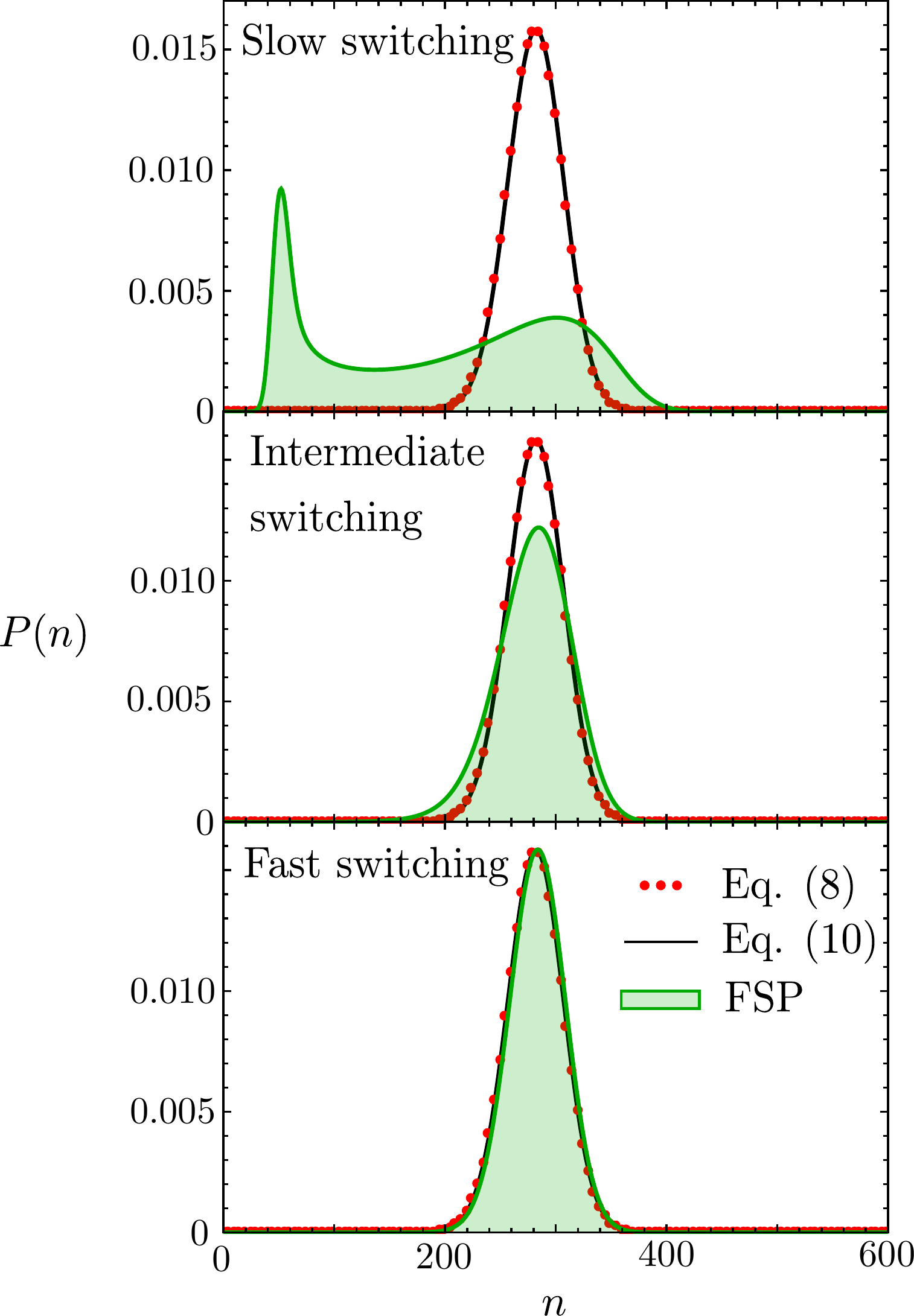}
\caption{Comparison of the heuristic reduced master equation solution from Eq. (\ref{eq:redMEsol}) (red dots), the FPE solution from Eq. (\ref{eq:FPsol1}) (black line) and the solution of the full cooperative network using FSP (green shaded region). Shared parameters in each plot are $r_u=50$, $r_b=400$, $\Omega=200$ and $d=1$. The FSP gives the exact solution for a truncated state space chosen such that the neglected probability mass is negligible. The top plot shows distributions for the case $s_u=s_b=5$, where clearly the heuristic master equation and FPE solutions are a poor approximation of the FSP. The middle plot shows distributions for the case $s_u=s_b=50$ where we can observe a convergence of the heuristic master equation and FPE solutions towards the FSP solution. The bottom plot shows excellent agreement of the FSP with the heuristic master equation and FPE solutions for fast switching where $s_u=s_b=5\times 10^3$.}
\label{fig0}
\end{figure}

The exact steady state solution of the one variable master equation given by Eq. (\ref{eq:redMEnb}) can be found using standard methods \cite{gardiner2009stochastic}:
\begin{align}\label{eq:redMEsol}
    P_a(n) = P_a(0)\prod_{z=1}^{n}\frac{T^{+}(z-1)}{T^{-}(z)},
\end{align}
where $P_a(0)$ is the steady state probability evaluated at $n = 0$ (acting effectively here as a normalisation constant).
We can further approximate the reduced master equation in Eq. (\ref{eq:redMEnb}) by the chemical FPE (via the Kramers-Moyal expansion) \cite{gillespie2000chemical,gardiner2009stochastic}:
\begin{align}\label{eq:FPE1}
    \frac{\partial P(n,t)}{\partial t} = -\frac{\partial}{\partial n}\big[a_1(n)P(n,t)\big]+\frac{1}{2}\frac{\partial^2}{\partial n^2}\big[a_2(n)P(n,t)\big],
\end{align}
where $a_1(n)$ and $a_2(n)$ are the first two jump moments, given by $a_1(n) = T^{+}(n)-T^{-}(n)$ and $a_2(n)=T^{+}(n)+T^{-}(n)$ respectively, and $P(n)$ denotes the FPE solution (a notation used throughout the paper). The purpose of this further approximation by means of a FPE will be made clear in Section \ref{sec:fdr}. Eq. (\ref{eq:FPE1}) has a steady state solution of the form \cite{van1992stochastic}:
\begin{align}\label{eq:FPsol1}
    P(n) = \frac{N}{T^{+}(n)+T^{-}(n)}\exp\Big(2\int^{n}\frac{T^{+}(z)-T^{-}(z)}{T^{+}(z)+T^{-}(z)}\,dz\Big),
\end{align}
where $N$ is a normalisation constant. Although the integral in the exponent of Eq. (\ref{eq:FPsol1}) can be solved exactly with propensities of the form of Eq. (\ref{eq:Propsnb}) and Eq. (\ref{eq:Propsnb2}) since it is the integral of the ratio of two cubic polynomials, the solution is too complicated to be detailed here. The approximations made by the FPE approximation are that (i) fluctuations in the protein number are small and (ii) we are in the fast switching regime between the gene states. Fig. \ref{fig0} compares the FPE solution Eq. (\ref{eq:FPsol1}) with the solution of the heuristic CME in Eq. (\ref{eq:redMEsol}) and the solution of the full CME of the reaction scheme in Eq. (\ref{eq:coopRS}) using the finite space projection method (FSP) \cite{munsky2006finite}. Note that provided the state space is truncated large enough, the FPE solution matches the solution of the heuristic CME almost exactly. Clearly, when gene switching is fast (bottom plot of Fig. \ref{fig0}) all three solutions agree with each other. However, when gene switching is not fast (top and middle plots on Fig. \ref{fig0}) both the reduced CME and FPE solutions are a poor approximation of the true distribution from FSP. 

\section{Accounting for fluctuating rates using the Unified Colored Noise Approximation}\label{sec3}

\begin{figure}[h!]
\centering
\includegraphics[width=0.8\textwidth]{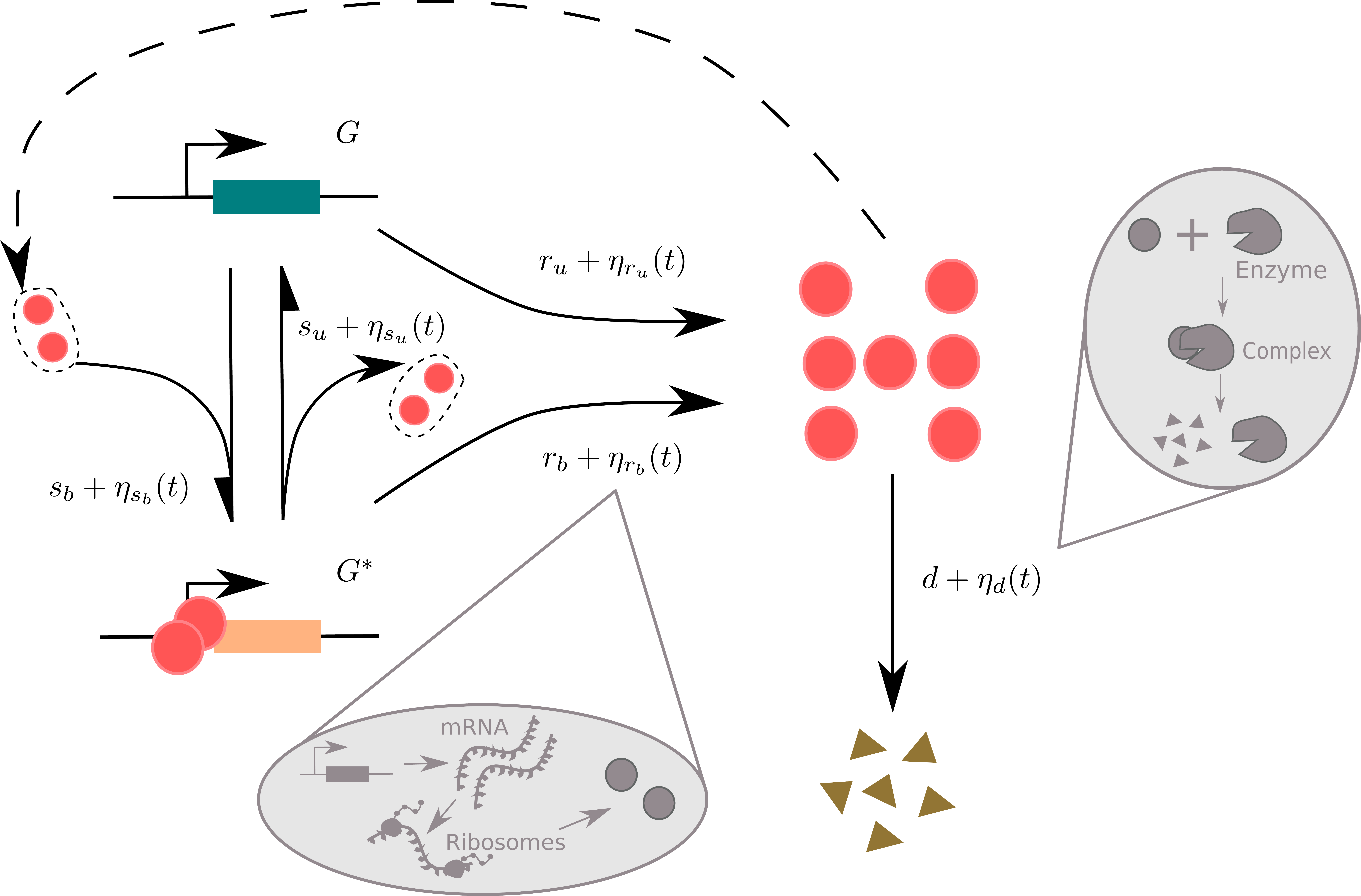}
\caption{Illustration of the cooperative auto-regulatory reaction scheme, with colored noise included on each individual reaction. For the case of non-fluctuating rates explored in Section \ref{sec2} the noise terms, $\eta_i$, on the rate parameter can be set to zero. Where colored noise is included in Section \ref{sec3} these noise terms are not set to zero. The addition of noise onto rate parameters can be thought of as accounting for processes that are not explicitly included in the gene expression model. Here we show two examples, where colored noise on the rate parameters of the reduced model can be used to account for mRNA number fluctuations during protein translation, or the degradation of proteins via an enzyme catalytic mechanism.}
\label{fig-cartoon}
\end{figure}

Fluctuating rate parameters can be used to include a description of processes not explicitly taken into account in the formulation of a model. In Fig. \ref{fig-cartoon} we illustrate this idea. In this section, we add fluctuations to the rate parameters of the FPE description derived earlier and use the UCNA to obtain a new effective FPE that is valid when the timescale of the noise on the rates is either very small or very large.

\subsection{Fluctuating degradation rate}\label{sec:fdr}
We begin by considering the case of a fluctuating degradation rate. These fluctuations could for example stem from details of the degradation machinery that are not explicitly described in the model, e.g multi-step degradation mediated by enzymatic reactions.

The equivalent Langevin equation to the chemical Fokker-Planck equation from Eq. (\ref{eq:FPE1}) using the propensities from Eqs. (\ref{eq:Propsnb}) and (\ref{eq:Propsnb2}) is given by \cite{gillespie2000chemical,van1992stochastic}:
\begin{align}\label{eq:equivLE1}
    \frac{d n}{dt} = \frac{r_u L +r_b (n/\Omega)^2}{L+(n/\Omega)^2}-d\,n + \sqrt{\frac{r_u L +r_b (n/\Omega)^2}{L+(n/\Omega)^2}+d\,n} \; \cdot \Gamma(t),
\end{align}
where $\Gamma(t)$ is Gaussian white noise with zero mean and correlator $\langle \Gamma(t) \Gamma(t') \rangle = \delta(t-t')$. Now we introduce a fluctuating degradation rate by setting $d = d_0(1+\eta(t))$, where $\eta(t)$ is Gaussian colored noise with a mean of zero and correlator $\langle \eta(t)\eta(t') \rangle = (D/\tau)\exp(-|t-t'|/\tau)$ \cite{jung1987dynamical,li1995bistable}. Here, $\tau$ is the \textit{correlation time} of the colored noise, $D/\tau$ is the \textit{noise strength} (the variance of fluctuations) and $d_0$ is the mean degradation rate. Since $D/\tau$ is the noise strength, i.e., $D$ scales the noise strength at constant $\tau$, we occasionally refer to $D$ itself as the \textit{noise strength} (where $\tau$ is a fixed parameter). In the limit of $\tau\to 0$ colored noise becomes white noise since $\lim_{\tau\to 0}\langle \eta(t)\eta(t') \rangle = D\delta(t-t')$. Note that $|\eta(t)|\ll1$ such that $d$ is a positive quantity (and hence admits physical interpretation as a rate parameter). The inclusion of colored noise can be shown to be equivalent to the following two component system \cite{jung1987dynamical}:
\begin{align}\label{eq:colLE1}
    \frac{d n}{dt} &= \frac{r_u L +r_b (n/\Omega)^2}{L+(n/\Omega)^2}-d_0(1+\eta)\,n + \sqrt{\frac{r_u L +r_b (n/\Omega)^2}{L+(n/\Omega)^2}+d_0\,n} \; \cdot \Gamma(t),\\\label{eq:colLE1b}
    \frac{d \eta}{dt} &= -\frac{1}{\tau}\eta + \frac{1}{\tau}\theta(t),
\end{align}
where $\theta(t)$ is Gaussian white noise with zero mean and correlator $\langle \theta(t) \theta(t') \rangle = 2 D \delta(t-t')$, and the time dependence on the protein number $n(t)$ and noise $\eta(t)$ is suppressed for notational convenience. Note that in the argument of the square root above we have replaced $\eta(t)$ by its mean of zero; this constitutes a mean-field type of approximation, and is necessary such that one can solve Eqs. (\ref{eq:colLE1})-(\ref{eq:colLE1b}) analytically -- however, where the noise is small this is generally a very good approximation. Note that we also use this mean field assumption in Sections \ref{sec:cnpp} and \ref{sec:cnbr}. For transparency, we rewrite Eqs. (\ref{eq:colLE1})-(\ref{eq:colLE1b}) as:
\begin{align}\label{eq:deg1}
    \frac{d n}{dt} &= h(n) + g_1(n) \eta + g_2(n) \Gamma(t),\\\label{eq:deg2}
    \frac{d \eta}{dt} &= -\frac{1}{\tau}\eta + \frac{1}{\tau}\theta(t),
\end{align}
with
\begin{align}
    h(n) &= \frac{r_u L +r_b (n/\Omega)^2}{L+(n/\Omega)^2}-d_0 n,\\
    g_1(n) &= -d_0 n,\\
    g_2(n) &= \sqrt{\frac{r_u L +r_b (n/\Omega)^2}{L+(n/\Omega)^2}+d_0\,n}.
\end{align}
In order to approximately solve Eqs. (\ref{eq:deg1})--(\ref{eq:deg2}) we next utilize the UCNA to obtain reduced Langevin equations when the noise $\eta$ is either very fast or very slow. For completeness, we present a non-rigorous but intuitive proof of the UCNA along the lines found in \cite{jung1987dynamical} which essentially consists of a direct adiabatic elimination on the stochastic differential equations (SDEs) in Eqs. (\ref{eq:deg1})--(\ref{eq:deg2}). For a more rigorous derivation of a UCNA-like FPE we advise reader to read the seminal work of Fox, who introduced a functional calculus approach to the study of colored noise SDEs \cite{fox1986functional,fox1986uniform,fox1987stochastic,fox1987steady}. A review of the differing UCNA-like derivations can be found in \cite{grigolini1988fokker}.

It has been discussed in \cite{jung1987dynamical,grigolini1988fokker} that the adiabatic elimination we employ below is exact for $\tau \to 0$ (white noise) or $\tau \to \infty$ (highly correlated noise) but that it should give a useful approximation for intermediate values of $\tau$. We note that the theory provided by Roberts et al. \cite{roberts2015dynamics} does not provide such a result as they consider separately the cases of $\tau\to0$ and $\tau\to\infty$. For us, the limit of $\tau\to\infty$ is not of biological interest, and we will later focus on the limit of $\tau$ small, although the derivation shown here holds for large $\tau$ too. First, where we use overdots to represent derivatives with respect to time $t$, one should proceed in rearranging Eq. (\ref{eq:deg1}) for $\eta$:
\begin{align}\label{eq:etadeg}
    \eta(n,\dot{n}) = \frac{1}{g_1(n)}(\dot{n}-h(n)-g_2(n) \Gamma(t)).
\end{align}
In what follows we will utilise a mean-field approximation (denoted by the subscript {\it{mf}}) to approximately calculate the time derivative of $\eta(n,\dot{n})$. We start by defining the mean-field approximation of $\eta(n,\dot{n})$ as:
\begin{align}
    \label{rgnew1}
    \eta_{mf}(n_{mf},\dot{n}_{mf}) = \frac{1}{g_1(n_{mf})}(\dot{n}_{mf}-h(n_{mf})).
\end{align}
Taking the time derivative with respect to non-dimensional time $\hat{t} = t/\tau$ (denoted by the overdot) we obtain:
\begin{align}\label{mfdot_A}
    \dot{\eta}_{mf}= \frac{1}{g_1(n_{mf})}\left(\frac{h(n_{mf})g_1'(n_{mf})}{g_1(n_{mf})}-h'(n_{mf})\right)\dot{n}_{mf}+\frac{\tau^{-1}}{g_1(n_{mf})}\left(\ddot{n}_{mf}-\frac{g_1'(n_{mf})}{g_1(n_{mf})}\dot{n}_{mf}^2\right),
\end{align}
where the prime on each function of $n_{mf}$ denotes the derivative with respect to $n_{mf}$. In the limit of $\tau \rightarrow 0$, the second term on the right hand side of Eq. \eqref{mfdot_A} goes to infinity and hence the only way to keep the time derivative finite is to impose the condition:
\begin{align}
    \ddot{n}_{mf}-\frac{g_1'(n_{mf})}{g_1(n_{mf})}\dot{n}_{mf}^2 = 0.
\end{align}
This then implies that in this limit we have:
\begin{align}\label{mfdotUC}
    \dot{\eta}_{mf} \approx \frac{1}{g_1(n_{mf})}\left(\frac{h(n_{mf})g_1'(n_{mf})}{g_1(n_{mf})}-h'(n_{mf})\right)\dot{n}_{mf}.
\end{align}
Note that taking the limit of $\tau \rightarrow \infty$ gives the same result and hence the approximation Eq. \eqref{mfdotUC} is valid in both the limit of small and large $\tau$.
This can be shown to be self-consistently true; taking the time-derivative of Eq. (\ref{eq:deg1}) alongside a mean-field approximation we get,
\begin{align}
    \ddot{n}_{mf} = \left(h'(n_{mf})+g'_1(n_{mf})\eta_{mf}\right)\dot{n}_{mf}+g_1(n_{mf})\dot{\eta}_{mf}.
\end{align}
Assuming Eqs. (\ref{rgnew1}) and (\ref{mfdotUC}) to be true one then recovers
\begin{align}\label{eq:degndd}
    \ddot{n}_{mf}-\frac{g_1'(n_{mf})}{g_1(n_{mf})}\dot{n}^2_{mf} = 0.
\end{align}

In Eq. (\ref{eq:deg2}) we can now substitute $\eta$ from (\ref{eq:etadeg}) and $\dot{\eta}_{mf}$ for $\dot{\eta}$ from Eq. (\ref{mfdotUC}) giving us the UCNA for the system with colored noise on the degradation rate, which is \textit{exact} in the limits $\tau\to 0$ or $\tau\to\infty$:
\begin{align}\label{eq:Cintro}
    \dot{n} \approx \frac{h(n)}{C(n,\tau)}+\frac{1}{C(n,\tau)}(g_1(n)\theta(t) + g_2(n) \Gamma(t)),
\end{align}
where
\begin{align}\label{eq:C}
    C(n,\tau) = 1+\tau\left(\frac{g'_1(n)h(n)}{g_1(n)}-h'(n) \right).
\end{align}
Note that we have dropped off the {\it{mf}} subscript for clarity. Finally, in order to get a simplified Langevin equation, we modify Eq. (\ref{eq:Cintro}) such that we only have one effective Gaussian white noise term. We begin by proposing:
\begin{align}\label{eq:totnoise}
    g(n)\Tilde{\Gamma}(t) = g_1(n)\theta(t) + g_2(n) \Gamma(t),
\end{align}
where $\Tilde{\Gamma}(t)$ is Gaussian white noise with mean zero and correlator $\langle \Tilde{\Gamma}(t)\Tilde{\Gamma}(t') \rangle = 2\delta(t-t')$, and then use relations between the correlators to find our unknown $g(n)$. Note that we assume zero correlation between $\Gamma(t)$ and $\theta(t)$, i.e. $\langle \Gamma(t)\theta(t')\rangle = \langle \Gamma(t')\theta(t)\rangle = 0$. Explicitly, utilising the correlators, we find:
\begin{align}
    g(n)^2\langle\Tilde{\Gamma}(t)\Tilde{\Gamma}(t')\rangle = g_1(n)^2\langle\theta(t)\theta(t')\rangle + g_2(n)^2\langle\Gamma(t)\Gamma(t')\rangle,
\end{align}
which gives us
\begin{align}
    g(n)=\sqrt{Dg_1(n)^2+\frac{1}{2}g_2(n)^2}.
\end{align}
Hence, our final reduced Langevin equation is given by:
\begin{align}\label{eq:UCNAdegLE}
    \dot{n} = \frac{h(n)}{C(n,\tau)}+\frac{g(n)}{C(n,\tau)}\Tilde{\Gamma}(t),
\end{align}
which corresponds to the result in \cite{li1995bistable}. Note that Eqs. (\ref{eq:Cintro}) and (\ref{eq:UCNAdegLE}) are \textit{identical}. Here we pause to make a couple of comments on $C(n,\tau)$, which can be interpreted as a renormalisation of the Langevin equation in Eq. (\ref{eq:deg1}) to account for the addition of colored noise to the rate parameters. In fact, when $\tau=0$, Eq. (\ref{eq:UCNAdegLE}) recovers the correct Langevin equation for a process with white noise on the rate parameters. One should also note the independence of $C(n,\tau)$ from the strength of the noise $D$; the renormalisation with respect to the addition of colored noise on the degradation rate is not specific to the size of the noise, it simply accounts for the finite correlation time.

The FPE corresponding to this SDE should be chosen in the Stratonovich form, following from \cite{fox1987stochastic,wong1965convergence,pesce2013stratonovich}, as this is the physical implementation of an SDE with colored noise having a non-zero correlation time $\tau$. This FPE is:
\begin{align}\label{eq:degFP}
    \frac{\partial P(n,t)}{\partial t} = -\frac{\partial}{\partial n}\left[\left(\Tilde{h}(n)+\Tilde{g}(n)\Tilde{g}'(n)\right)P(n,t)\right] + \frac{\partial^2}{\partial n^2}\left[\Tilde{g}(n)^2P(n,t) \right],
\end{align}
where $\Tilde{h}(n) = h(n)/C(n,\tau)$ and $\Tilde{g}(n) = g(n)/C(n,\tau)$. Following Eqs. (\ref{eq:FPE1})--(\ref{eq:FPsol1}) in Section \ref{sec2} and \cite{van1992stochastic}, the steady state solution to this equation is then given by:
\begin{align}\label{eq:UCNAdeg}
    P(n) = \frac{N}{\Tilde{g}(n)^2}\exp\left(\int^n \frac{\Tilde{h}(z)+\Tilde{g}(z)\Tilde{g}'(z)}{\Tilde{g}(z)^2} dz \right)=\frac{N}{\Tilde{g}(n)}\exp\left(\int^n \frac{\Tilde{h}(z)}{\Tilde{g}(z)^2} dz \right),
\end{align}
where $N$ is the normalisation constant, chosen over the domain $n\in [0,\infty)$.


To test the accuracy of the distributions for colored noise provided by the UCNA in Eq. (\ref{eq:UCNAdeg}), we compare the UCNA solution to a distribution produced from a modified SSA that explicitly accounts for the colored noise on the degradation rate. This modification is given in full detail in Appendix \ref{sec:ssa}. Essentially, the dilution/degradation reaction $P\to\varnothing$ is replaced by three new reactions alongside the introduction of a ghost species $Y$, these being (i) $\varnothing\xrightleftharpoons[]{}Y$ and (ii) $P+Y\to Y$. The rates of these new reactions are then chosen to ensure the magnitude of \textit{effective} external noise on the degradation reaction, due to fluctuations in molecule numbers of the ghost species, match the colored noise SDE given in Eq. (\ref{eq:colLE1b}).

\begin{figure}[h!]
\centering
\includegraphics[width=1.0\textwidth]{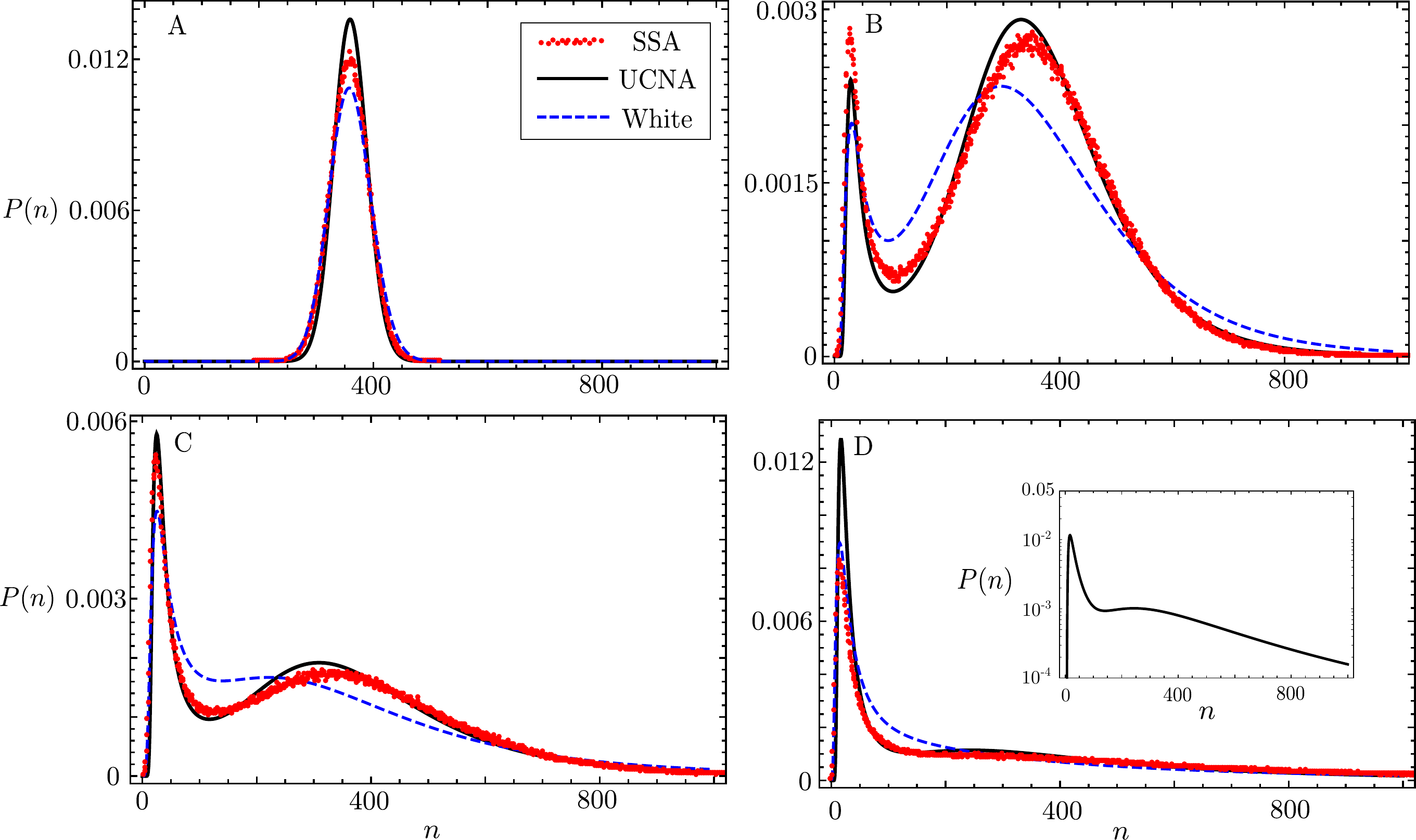}
\caption{Comparison of the UCNA (black line) from Eq. (\ref{eq:UCNAdeg}) and white extrinsic noise (UCNA with $\tau=0$, dashed blue line) with stochastic simulations using the modified SSA (red  points) of the cooperative reaction scheme in Eq. (\ref{eq:coopRS}), where the colored noise is added to the degradation rate. Aside from variation in the strength of noise $D$ (shown on each plot), the shared parameters are $r_u=24,\,r_b=464,\,s_b=s_u=1000,\,d_0=1,\Omega=200$ and $\tau=1$. Parameters $s_b$ and $s_u$ are chosen to be large compared to other system parameters such that the frequency of gene activation and inactivation events is much larger than the frequency of other reaction events, i.e. the fast gene switching assumption. Note that for this choice of rate parameters, the rate equations are bistable with equilibrium points at $n=47.4,360.4$. The criterion $\sqrt{D/\tau}<1$ is required to ensure positivity of the degradation rate. As the extrinsic noise is increased, the mass of the distribution shifts from the mode at 360.4 to the mode at 47.4. The inset of D shows the same distribution but with the y-axis on a log scale, emphasising the exponential tail of the distribution for large $n$. SSA data in each case comes from a single steady state trajectory of $9\times 10^6$s.}
\label{fig1}
\end{figure} 

Fig. \ref{fig1} shows steady state probability distributions produced by the UCNA for various values of $D$ for a deterministically bistable set of parameters. The UCNA correctly captures the shift of the probability mass from the equilibrium point of higher molecule number (referred to as the \textit{upper mode}) to the lower equilibrium point (referred to as the \textit{lower mode}) as $D$ is increased. Importantly, this shows that when gene switching is assumed to be fast, colored noise can induce bimodality -- one should keep this in mind for when we look at slow gene switching in Section \ref{cUCNA}. Readers should also note that the parameter choices have been selected such that the Fokker-Planck approximation is good, notably that the system size is large, i.e., $\Omega \gg 1$, and the mean number of proteins in the system is also large. In all cases $\sqrt{D/\tau}<1$ so that the degradation rate remains positive. The behaviour seen as $D$ increases in Fig. \ref{fig1} can be explained as follows. When $D$ is small (Fig. \ref{fig1}A) the colored noise $\eta$ in Eq. (\ref{eq:deg2}) is also small compared to the mean number of molecules in the system, and the noise cannot force the system out of the upper mode. As $D$ gets larger (Figs. \ref{fig1}B and \ref{fig1}C) the fluctuations $\eta$ at the upper mode also become larger, allowing the system to explore the lower mode. When the system is found in the lower mode the pre-factor of the coloured noise in Eq. (\ref{eq:deg1}), $g_1(n) \propto n$, is lesser in magnitude, and the fluctuations in $\eta$ are much smaller than when the system inhabits the upper mode hence the increased probability mass at the lower mode. That the system is less noisy at the lower mode means that the system struggles more to get a fluctuation large enough to propel it into the upper mode. These properties of the system as $D$ increases can be further seen through (i) the increase in probability mass found at the lower mode as $D$ increases thoroughout all of Fig \ref{fig1}(A--D), and (ii) the increased probability mass found in the tail of the distribution for large $n$ (Fig. \ref{fig1}D); while the tail is very slowly decaying it is still exponential and hence the distribution is not heavy-tailed (see the inset of Fig. \ref{fig1}D). This ability to induce bimodality through a more detailed description of the details of the degradation process is important in the context of \textit{cellular decision-making}. It is hence possible for regions of the reaction rate parameter space previously thought unable to induce multiple phenotypic states to do so with an increasing influence of more complex degradation mechanisms. Note that for the majority of cases in Fig. \ref{fig1}, the UCNA provides a much better approximation than the white noise approximation, hence one cannot simply assume that since the correlation time $\tau$ is relatively small that it can be approximated as zero.

\begin{figure}[h!]
\centering
\includegraphics[width=1.0\textwidth]{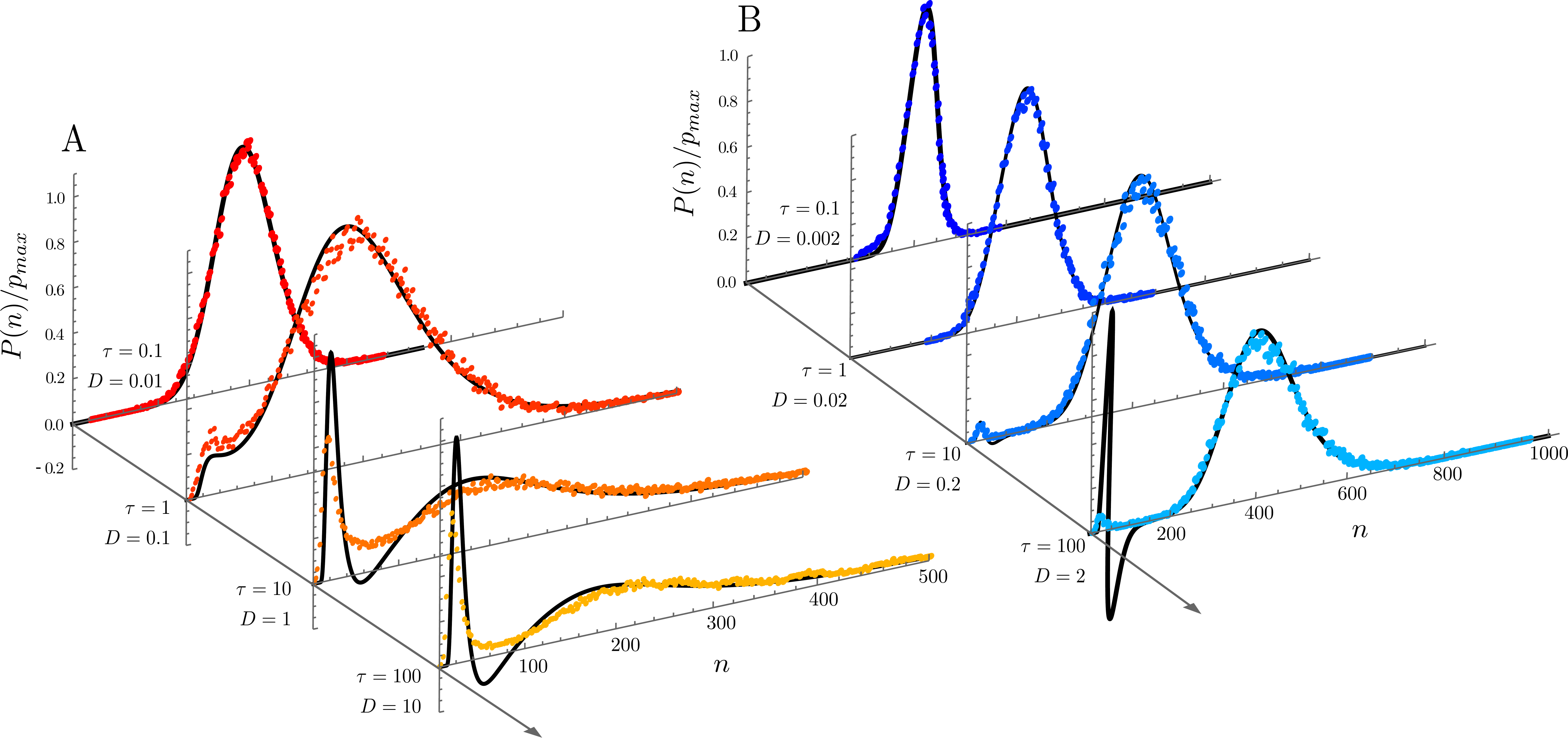}
\caption{Comparison of the UCNA (black line) against the modified SSA (colored dots) as the correlation time $\tau$ is increased at constant noise size $D/\tau$. Note that the y-axis shows $P(n)/p_{max}$, where $P(n)$ is defined in Eq. (\ref{eq:UCNAdeg}) and $p_{max}$ is equal to the maximum value of $P(n)$. (A) Shows the performance of increasing $\tau$ for a system with parameters $r_u=20,\,r_b=250,\,d_0=1,\,s_u=3\times10^2,\,s_b=10^3$ and $\Omega=100$. Deterministically this system is monostable with an equilibrium point at $n=194.7$, however as $\tau$ is increased a shift towards a lower mode is observed. When $\tau$ is sufficiently large, the UCNA predicts a negative probability. (B) Shows similar to (A) but with parameters $r_u=25,\,r_b=480,\,d_0=1,\,s_u=8\times10^2,\,s_b=10^3$ and $\Omega=200$. This too is a deterministically monostable system with equilibrium point $n=406.0$. As $\tau$ increases, the breakdown of the UCNA is more apparent than for (A) with the prediction of negative probability for small $n$ more drastic. Both (A) and (B) show that unless $\tau$ is large, while $D/\tau$ is small, the UCNA provides a very good approximation, even where the colored noise induces bimodality in deterministically monomodal systems. SSA data in each case comes from a single steady state trajectory of $9\times 10^5$s.}
\label{fig2}
\end{figure} 

Fig. \ref{fig2} shows how the UCNA responds to increasing correlation time $\tau$ while the noise strength, $D/\tau$, remains fixed. For all cases where $\tau$ is small, the UCNA performs very well. As $\tau$ increases however the UCNA starts to predict ever increasing negative probabilities for some values of $n$.
Notably though, Fig. \ref{fig2}B shows that even where significant negative probability is predicted at large $\tau$, the UCNA still manages to capture the rest of the distribution. This negativity of $C(n,\tau)$ is commented on in  both \cite{jung1987dynamical} and \cite{fox1987stochastic}. The former deals with this negativity by taking the absolute magnitude of the pre-factor of the exponential in Eq. (\ref{eq:UCNAdeg}), while the latter comments that the proof of their UCNA-like FPE is only formally valid where $C(n,\tau)>0,\;\forall \;n$. Here we choose not to take the magnitude of the pre-factor in Eq. (\ref{eq:UCNAdeg}), since although this leads to a positive probability for all $n$ it is nonetheless a poor approximation; but we take careful note of the comment made by Fox in \cite{fox1987stochastic}, as this indicates where the UCNA will perform well. The intuition behind the argument of Fox can be stated as: if for some $n$, $C(n,\tau)<0$ there must be a transitory value of $n$ for which $C(n,\tau)=0$, at this point the Eq. (\ref{eq:UCNAdegLE}) becomes physically ill-defined and our solution is invalid.

Finally, we observe that the parameter values chosen for both plots in Fig. \ref{fig2} correspond to deterministically monostable systems. The bimodality that is observed in Fig. \ref{fig2} is hence \textit{noise induced bimodality}. The mode that appears for small $\tau$ corresponds to the deterministic equilibrium point, whereas the noise induced mode does not correspond to an equilibrium point of the deterministic system. We notice that the ability to exhibit a noise induced mode as $\tau$ becomes large is especially true for monostable parameter sets which are in close proximity to bistable parameter sets in the parameter space. This can be explained by occasional jumps between the monostable and bistable regimes due to sufficiently large fluctuations in the degradation rate. Hence a measure of the distance here is the difference in the magnitude of $d_0$ needed such that the system is deterministically bistable divided by the noise strength, defined as $\Delta d_0 = |d_0-d_c|/(D/\tau)$, where $d_c$ is the closest value of the mean degradation rate to $d_0$ expressing bistability. For example, the parameter set chosen in Fig. \ref{fig2}A, although monostable, is very close to a parameter set that exhibits deterministic bistability ($\Delta d_0 = 2.12$). On the other hand, the parameter set of Fig. \ref{fig2}B is far from the bistable parameter regime ($\Delta d_0 = 57.5$) -- and hence the bimodality shown is very limited as $\tau$ becomes large. The reason for this noise induced bimodality then can be seen by the ability of a system, through fluctuations in the rate parameters, to access parameter regimes which in fact do exhibit deterministic bistability. Importantly, even when it seems bimodality is not induced (e.g., Figs. \ref{fig1}A or \ref{fig2}A), using the extremal equation of $P(n)$ from \cite{li1995bistable}, i.e., $\Tilde{h}(n)=\Tilde{g}(n)\Tilde{g}'(n)$, one can show that the UCNA still predicts the presence of two modes. This explanation of the induced bimodality in cooperative autoregulation is further supported by the lack of noise induced bimodality when colored noise is included on the degradation rate of the FPE describing non-cooperative autoregulation; here the UCNA's extremal equation only ever predicts the existence of one mode for the probability distribution.



\subsection{Fluctuating effective protein production rates}\label{sec:cnpp}

We now extend the analysis from Section \ref{sec:fdr} to the effective protein production rates. Colored noise on the effective production rates can be used to implicitly model multi-step protein production, including multiple stages of mRNA processing before translation (see Fig. \ref{fig-cartoon}). We add colored noise onto the effective protein production rates via, $r_u = r_u^{(0)}(1+\eta_1(t))$ and $r_b = r_b^{(0)}(1+\eta_2(t))$, which upon substituting in the Langevin equation describing the feedback loop Eq. \eqref{eq:equivLE1} we obtain the following set of SDEs:
\begin{align}\label{eq:colLE2}
    \frac{d n}{dt} &= \frac{r_u^{(0)} L +r_b^{(0)} (n/\Omega)^2}{L+(n/\Omega)^2}-d\,n + \frac{r_u^{(0)}L\eta_1 +r_b^{(0)} (n/\Omega)^2\eta_2}{L+(n/\Omega)^2} + \sqrt{\frac{r_u^{(0)} L +r_b^{(0)} (n/\Omega)^2}{L+(n/\Omega)^2}+d\,n} \; \cdot \Gamma(t),\\
    \frac{d \eta_1}{dt} &= -\frac{1}{\tau}\eta_1 + \frac{1}{\tau}\theta_1(t),\\
    \frac{d \eta_2}{dt} &= -\frac{1}{\tau}\eta_2 + \frac{1}{\tau}\theta_2(t),
\end{align}
where $\theta_1(t)$ and $\theta_2(t)$ are Gaussian white noise terms with zero mean and correlators $\langle \theta_1(t) \theta_1(t') \rangle = 2 D_1 \delta(t-t')$ and $\langle \theta_2(t) \theta_2(t') \rangle = 2 D_2 \delta(t-t')$ respectively. Note that here we have used a mean field approximation for the terms under the square root, as was done in Section \ref{sec:fdr}. In a similar style to Eq. (\ref{eq:totnoise}) we now propose a new noise term $\Tilde{\eta}(t)$, which couples $\eta_1(t)$ and $\eta_2(t)$, satisfying:
\begin{align}
    F(n)\Tilde{\eta}(t) = f_1(n)\eta_1(t)+f_2(n)\eta_2(t),
\end{align}
where $f_1(n) = r_u^{(0)}L/(L+(n/\Omega)^2)$, $f_2(n) = r_b^{(0)}(n/\Omega)^2/(L+(n/\Omega)^2)$ and $\Tilde{\eta}(t)$ is colored noise with zero mean and correlator $\langle \Tilde{\eta}(t)\Tilde{\eta}(t')\rangle = e^{-|t-t'|/\tau}/\tau$, satisfying the following equation:
\begin{align}\label{eq:etatil}
\frac{d \Tilde{\eta}}{dt} = -\frac{1}{\tau}\Tilde{\eta} + \frac{1}{\tau}\theta(t),
\end{align}
where $\theta(t)$ is Gaussian white noise with correlator $\langle \theta(t)\theta(t') \rangle = 2\delta(t-t')$. The correlators for $\eta_1(t)$ and $\eta_2(t)$ are $\langle \eta_1(t)\eta_1(t')\rangle = D_1 e^{-|t-t'|/\tau}/\tau$ and  $\langle \eta_2(t)\eta_2(t')\rangle = D_2 e^{-|t-t'|/\tau}/\tau$, where we have assumed that the colored noise on both production rates has the same correlation time but a differing magnitude of noise strength. Using the properties of the correlators of $\eta_1$, $\eta_2$ and $\Tilde{\eta}$ we then find:
\begin{align}
    F(n) = \sqrt{f_1(n)^2D_1+f_2(n)^2D_2}.
\end{align}
Sharing the notation adopted in Section \ref{sec:fdr}, we define the following:
\begin{align}
     h(n) &= \frac{r_u^{(0)} L +r_b^{(0)} (n/\Omega)^2}{L+(n/\Omega)^2}-d n,\\
     g_2(n) &= \sqrt{\frac{r_u^{(0)}L+r_b^{(0)}(n/\Omega)^2}{L+(n/\Omega)^2}+d n}.
\end{align}
This gives us the following SDE which is coupled to Eq. (\ref{eq:etatil}): 
\begin{align}\label{midwaySDE}
    \frac{dn}{dt} = h(n) + F(n)\Tilde{\eta}+ g_2(n)\Gamma(t).
\end{align}
Then, following the same UCNA procedure as in Eqs. (\ref{eq:etadeg})--(\ref{eq:Cintro}), we obtain the following approximate Langevin equation:
\begin{align}\label{eq:eppPreLE}
     \dot{n} \approx \frac{h(n)}{C(n,\tau)}+\frac{1}{C(n,\tau)}(F(n)\theta(t)+g_2(n)\Gamma(t)),
\end{align}
where
\begin{align}
    C(n,\tau) = 1+\tau\left(\frac{F'(n)h(n)}{F(n)}-h'(n)\right).
\end{align}
In this case it is interesting to note that unlike the case of a fluctuating degradation rate, here $C(n,\tau)$ does depend on both the correlation time $\tau$ and the strength of the colored noise $D_1,D_2$ (unless $D_1 = D_2$ in which case there is only dependence on $\tau$). To simplify Eq. (\ref{eq:eppPreLE}) further, we again propose:
\begin{align}
    g(n)\Tilde{\Gamma}(t) = F(n)\theta(t) + g_2(n) \Gamma(t),   
\end{align}
where $\Tilde{\Gamma}(t)\Tilde{\Gamma}(t')=2\delta(t-t')$, and find using the correlators that $g(n) = \sqrt{F(n)^2+g_2(n)^2/2}$. This leads to the final approximate SDE:
\begin{align}\label{eq:UCNAtransLE}
    \dot{n} = \frac{h(n)}{C(n,\tau)}+\frac{g(n)}{C(n,\tau)}\Tilde{\Gamma}(t),
\end{align}
which is identical in notation to Eq. (\ref{eq:UCNAdegLE}) but where $h(n)$, $C(n,\tau)$ and $g(n)$ are all defined in this section. The equivalent FPE for this SDE is then:
\begin{align}\label{eq:prodFP}
    \frac{\partial P(n,t)}{\partial t} = -\frac{\partial}{\partial n}\left[\left(\Tilde{h}(n)+\Tilde{g}(n)\Tilde{g}'(n)\right)P(n,t)\right] + \frac{\partial^2}{\partial n^2}\left[\Tilde{g}(n)^2P(n,t) \right].
\end{align}
Again our solution for the probability distribution will then be:
\begin{align}\label{CNprodProb}
    P(n)=\frac{N}{\Tilde{g}(n)}\exp\left(\int^n \frac{\Tilde{h}(z)}{\Tilde{g}(z)^2} dz \right),
\end{align}
with $\Tilde{h}(n)=h(n)/C(n,\tau)$ and $\Tilde{g}(n)=g(n)/C(n,\tau)$.
\begin{figure}[h!]
\centering
\includegraphics[width=1.0\textwidth]{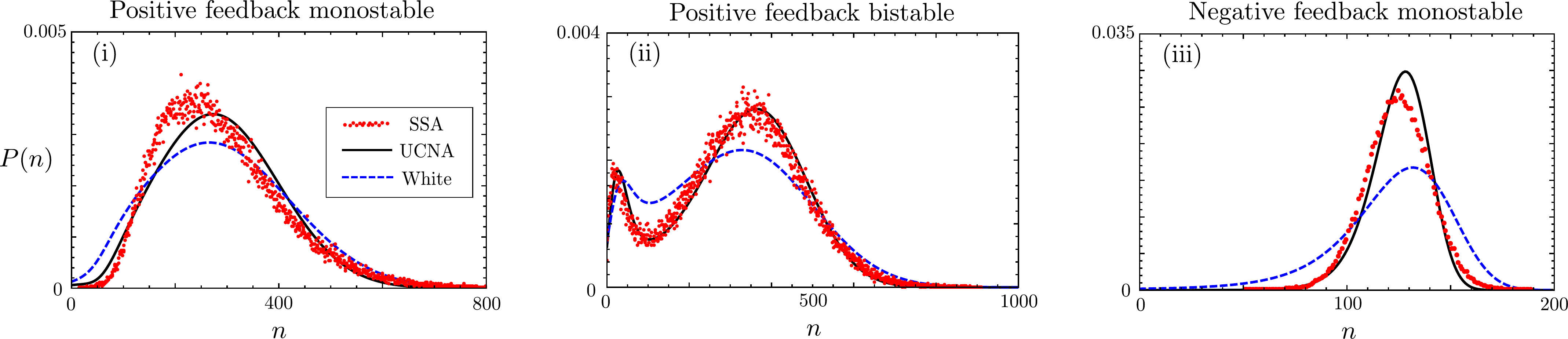}
\caption{This figure shows the agreement of the UCNA on the protein production rates to the modified SSA (detailed in Section \ref{sec:cnpp}), also compared to the case of white extrinsic noise ($\tau=0$). The plots show agreement of the UCNA over the three main qualitative regimes of cooperative autoregulation at large molecule number, showing respectively: (i) monostable positive feedback with parameters $r_u^{(0)}=150$, $r_b^{(0)}=300$, $s_u=s_b=10^3$, $d=1$, $D_1=0.25$, $D_2=0.25$, $\tau=0.5$, $\Omega=100$; (ii) bistable positive feedback with parameters $r_u^{(0)}=24$, $r_b^{(0)}=468$, $s_u=s_b=10^3$, $d=1$, $D_1=0.75$, $D_2=0.1$, $\tau=1$, $\Omega=200$; (iii) monostable negative feedback with parameters $r_u^{(0)}=470$, $r_b^{(0)}=20$, $s_u=s_b=10^3$, $d=1$, $D_1=0.1$, $D_2=0.1$, $\tau=1$, $\Omega=70$. SSA data in each case comes from a single steady state trajectory of $9\times 10^5$s.}
\label{fig5}
\end{figure} 

We now describe the modified SSA that takes into account extrinsic noise on the effective protein production rates. This modification replaces the protein production reaction in each gene state, i.e., $G_k\to G_k+P$ where $G_k$ represents either $G$ or $G^*$, by three new reactions alongside the introduction of a ghost species $Y_k$ for each gene state. These new reactions are $\varnothing\xrightleftharpoons[r_2]{r_1}Y_k$ and $G_k+Y_k\xrightarrow{r_3} G_k + Y_k + P$. Utilising the LNA (assuming $Y_k$ to be abundant), as was done for colored noise on the degradation rate in Appendix \ref{sec:ssa}, one finds these rates to be $r_1 = 1/(D_k \Omega)$, $r_2 = 1/\tau$ and $r_3 = r_k^{(0)}D_k \Omega/\tau$,  which ensure matching to the colored noise SDE given in Eq. (\ref{eq:colLE2}), where $r_k^{(0)}$ represents $r_u^{(0)}$ or $r_b^{(0)}$ in $G$ and $G^*$ respectively.

Figure \ref{fig5} shows a good performance of the UCNA when compared to the modified SSA described above. This performance is shown for each differing qualitative behaviour expressed by cooperative bimodality, i.e., (i) monostable positive feedback, (ii) bistable positive feedback, and (iii) monostable negative feedback. In all three plots shown the UCNA matches the modified SSA well, and clearly performs better than if one were to approximate the colored noise with white noise (i.e., $\tau=0$).

We find the same qualitative behaviour of the creation and eventual destruction of bimodality (see Fig. \ref{fig6}A(i--iii)) as the noise strengths, $D_1$ and $D_2$, become large for the colored noise on the protein production rates as was found in Fig. \ref{fig1} for  colored noise on the degradation rate. Note that for the chosen parameter set in Fig. \ref{fig6}A that the white noise approximation performs generally very well compared to the UCNA. For $\tau\leq1$, the white extrinsic noise approximation can typically perform quite well compared to the modified SSA, but note that this is not always the case (e.g., see Fig. \ref{fig1}).
\subsection{Fluctuating binding/unbinding rates}\label{sec:cnbr}
Finally, we apply the UCNA to the case of colored noise added to the binding and unbinding rates of the protein to the gene. This could be utilised to implicitly model the effect of multiple gene states in the transition of $G$ to $G^*$, as has been experimentally and theoretically investigated \cite{nicolas2017shapes,cao2019multi,zhang2019stationary}, accounting for DNA looping via distal enhancers or chromatin conformational states. For convenience we define $s_b=s_b^{(0)}(1+\eta_1(t))$, $s_u=s_u^{(0)}(1+\eta_2(t))$ and
\begin{align}\label{Lapp}
    L_\eta = L_0 \left(\frac{1+\eta_1(t)}{1+\eta_2(t)}\right),\;\text{with }\; L_0 = \frac{s_u^{(0)}}{s_b^{(0)}}.
\end{align}
Substituting Eq. (\ref{Lapp}) in the Langevin equation describing the feedback loop Eq. \eqref{eq:equivLE1} (and making a mean-field approximation for the terms under the square root) we obtain the following set of SDEs:
\begin{align}\label{eq:bindingNonLin}
\frac{d n}{dt} &= \frac{L_\eta r_u + r_b(n/\Omega)^2}{L_\eta+(n/\Omega)^2} + \sqrt{\frac{r_u L_0 +r_b(n/\Omega)^2}{L_0+(n/\Omega)^2}+d\,n} \; \cdot \Gamma(t),\\
    \frac{d \eta_1}{dt} &= -\frac{1}{\tau}\eta_1 + \frac{1}{\tau}\theta_1(t),\\
    \frac{d \eta_2}{dt} &= -\frac{1}{\tau}\eta_2 + \frac{1}{\tau}\theta_2(t),
\end{align}
where $\theta_1(t)$ and $\theta_2(t)$ are Gaussian white noise terms with zero mean and correlators $\langle \theta_1(t) \theta_1(t') \rangle = 2 D_1 \delta(t-t')$ and $\langle \theta_2(t) \theta_2(t') \rangle = 2 D_2 \delta(t-t')$ respectively. In order to proceed using the UCNA we must linearise the drift term in Eq. (\ref{eq:bindingNonLin}) with respect to $\eta_1$ and $\eta_2$ through the small noise approximation $\eta_1$,$\eta_2 \ll 1$:
\begin{align}\label{leeqsmn}
    \frac{d n}{d t}\approx \frac{r_u L_0 + r_b (n/\Omega)^2}{L_0 + (n/\Omega)^2}- d\, n+ \left(\frac{L_0 n^2 \Omega ^2 \left(r_u-r_b\right)}{\left(L_0 \Omega ^2+n^2\right){}^2}\right)(\eta_1-\eta_2)\\\nonumber+\sqrt{\frac{r_u L_0 +r_b (n/\Omega)^2}{L_0+(n/\Omega)^2}+d\,n} \; \cdot \Gamma(t).
\end{align} 
\begin{figure}
\centering
\includegraphics[width=1.0\textwidth]{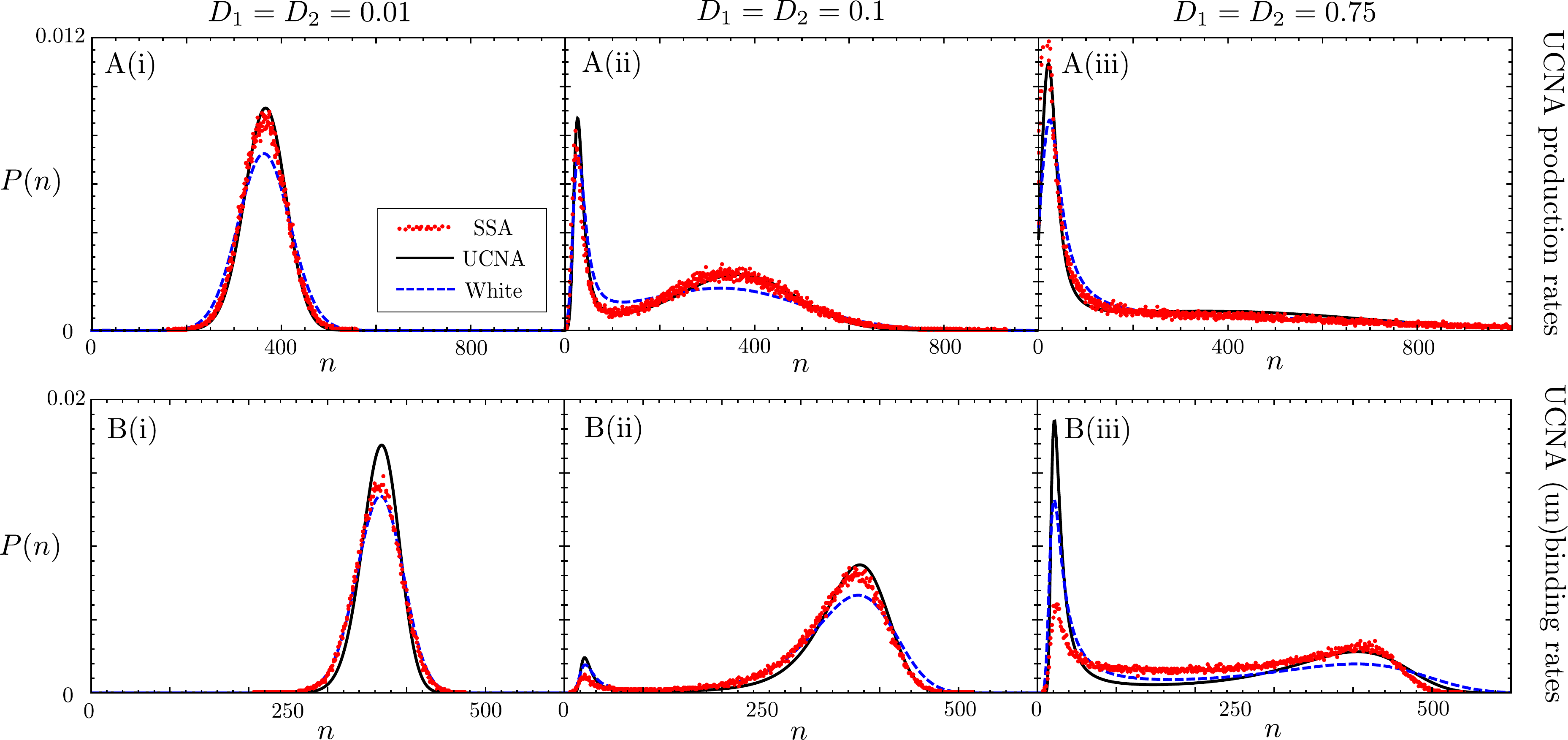}
\caption{Plots showing the creation and eventual destruction of bimodality in the probability distributions for colored noise on the (A) protein production rates, (B) binding/unbinding rates (denoted on the figure by (un)binding rates), analogously to what was observed in Fig. \ref{fig1} for colored noise on the degradation rate. For A it is clear that the UCNA performs well where the noise strength is both small in A(i) and large in A(iii). For B we see that the low (B(i)) and intermediate (B(ii)) noise cases are well predicted by the UCNA and white noise approximation, however where the noise becomes large (B(iii)) the UCNA breaks down, whereas the white noise approximation still performs well compared to the modified SSA prediction. Other than the noise strengths given on the figure, the parameters for both A and B are $r_u^{(0)}=24$, $r_b^{(0)}=468$, $s_u=s_b=10^3$, $d=1$, $\tau=1$, $\Omega=200$ (i.e., the same parameters used in Fig. \ref{fig1} which express deterministic bistability). SSA data in each case comes from a single steady state trajectory of $9\times 10^5$s.}
\label{fig6}
\end{figure} 
For convenience we now define:
\begin{align}
    h(n) &= \frac{r_u L_0 + r_b (n/\Omega)^2}{L_0 + (n/\Omega)^2}- d\, n,\\
    g_1(n) &= \frac{L_0 n^2 \Omega ^2 \left(r_u-r_b\right)}{\left(L_0 \Omega ^2+n^2\right){}^2},\\
    g_2(n) &= \sqrt{\frac{r_u L_0 +r_b (n/\Omega)^2}{L_0+(n/\Omega)^2}+d\,n},\\\label{FforBR}
    F(n) &= g_1(n)\sqrt{D_1+D_2},\\
    g(n) &= \sqrt{F(n)^2+g_2(n)^2/2}.
\end{align}
In terms of these new functions Eq. (\ref{leeqsmn}) becomes,
\begin{align}
    \frac{d n}{d t} = h(n)+g_1(n)(\eta_1-\eta_2)+g_2(n)\Gamma(t).
\end{align}
Following Section \ref{sec:cnpp} we then arrive at the UCNA for colored noise on the binding rates where $\eta_1$,$\eta_2 \ll 1$:
\begin{align}
    \frac{d n}{d t} = \frac{h(n)}{C(n,\tau)} + \frac{1}{C(n,\tau)}\left(F(n)\theta(t)+g_2(n)\Gamma(t) \right).
\end{align}
Then, using the properties of the correlators of $\theta(t)$ and $\Gamma(t)$ we arrive at:
\begin{align}\label{eq:UCNA-Lord1}
    \frac{d n}{d t} = \frac{h(n)}{C(n,\tau)}+\frac{g(n)}{C(n,\tau)}\Tilde{\Gamma}(t),
\end{align}
where,
\begin{align}\label{CforBR}
    C(n,\tau) = 1+\tau\left(\frac{g_1'(n)h(n)}{g_1(n)}-h'(n)\right),
\end{align}
and $\Tilde{\Gamma}(t)$ is Gaussian white noise with mean zero and correlator $\langle \Tilde{\Gamma}(t)\Tilde{\Gamma}(t')\rangle = 2\delta(t-t')$. Here, as for the UCNA applied to the degradation rate, $C(n,\tau)$ is again independent of the strengths of the colored noise terms. This UCNA, as we shall see, should be a good approximation where both $D_1$ and $D_2$ are small -- by `small' we explicitly mean that $D_1$ and $D_2$ should be smaller than noise strengths used on the UCNA for protein production rates or the degradation rate. The solution to Eq. (\ref{eq:UCNA-Lord1}) is given by:
\begin{align}
    P(n)=\frac{N}{\Tilde{g}(n)}\exp\left(\int^n \frac{\Tilde{h}(z)}{\Tilde{g}(z)^2} dz \right),
\end{align}
with $\Tilde{h}(n)=h(n)/C(n,\tau)$ and $\Tilde{g}(n)=g(n)/C(n,\tau)$.

Now we evaluate the performance of the UCNA on the binding and unbinding rates, and compare it with the modified SSA. In this case the modified SSA replaces the binding and unbinding reactions, $G+2P\xrightleftharpoons[s_u]{s_b}G^*$, by the following: $\varnothing\xrightleftharpoons[r_2]{r_1}Y_1$, $G+Y_1+2P\xrightarrow{r_3}Y_1+G^*$, $\varnothing\xrightleftharpoons[r_5]{r_4}Y_2$, and $G^*+Y_2 \xrightarrow{r_6}G+Y_2+2P$, where $Y_1$ and $Y_2$ are ghost species. The rates of these reactions are determined via the LNA (assuming the ghost species to be numerous) and are $r_1 = 1/(D_1\Omega)$, $r_2 = 1/\tau$, $r_3 = s_b^{(0)}D_1\Omega/\tau$, $r_4 = 1/(D_2\Omega)$, $r_5 = 1/\tau$ and $r_6 = s_u^{(0)}D_2\Omega/\tau$.

In Fig. \ref{fig6}B we test the UCNA on the binding and unbinding rates compared to the modified SSA described above. Clearly, the same qualitative behaviour of the creation and destruction of bimodality, as noise strength is increased, is observed, as was also observed for colored noise on the degradation rate (Fig. \ref{fig1}) and protein production rates (Fig. \ref{fig6}A). The resultant expression of bimodality however, is clearly different than for these cases. Notably, this UCNA does ascribe to an additional limitation compared to the UCNA of degradation or production rates; a limitation due to the further small noise approximation made in Eq. (\ref{leeqsmn}). This limitation is seen in Fig. \ref{fig6}B(iii), showing that the UCNA applied to the binding and unbinding rates is much more sensitive to increased noise strength than the other UCNA applications. One also observes that the white noise approximation in Fig. \ref{fig6}B performs  almost as well as the UCNA (Figs. \ref{fig6}B(i--ii)) or in some cases better than the UCNA (Fig. \ref{fig6}B(iii)); hence, the white noise approximation may be a safer approximation than the UCNA for colored noise applied to the binding and unbinding rates.

\subsection{Breakdown conditions of the UCNA}\label{breakdown}

Having now applied the UCNA to approximate distributions for colored noise on the (i) degradation rate, (ii) protein production rates and (iii) the binding/unbinding rates, we now assess the conditions which cause the UCNA to breakdown. The application of the UCNA to colored noise on the protein production rates presents a somewhat more complex problem than the application of the UCNA to colored noise on the degradation rate or the binding/unbinding rates; hence, we more easily see that there are \textit{three main conditions for the breakdown of the UCNA} -- conditions beside the large system size or large molecule number requirement needed to approximate the discrete master equation by a one variable FPE. Below we detail these three conditions, in each case explaining why the disagreement occurs. Note that although the analysis of breakdown conditions below is done for the UCNA on the protein production rates, the same arguments hold for the other applications of the UCNA previously presented.

\subsubsection{Condition 1}
\begin{figure}[h!]
\centering
\includegraphics[width=1.0\textwidth]{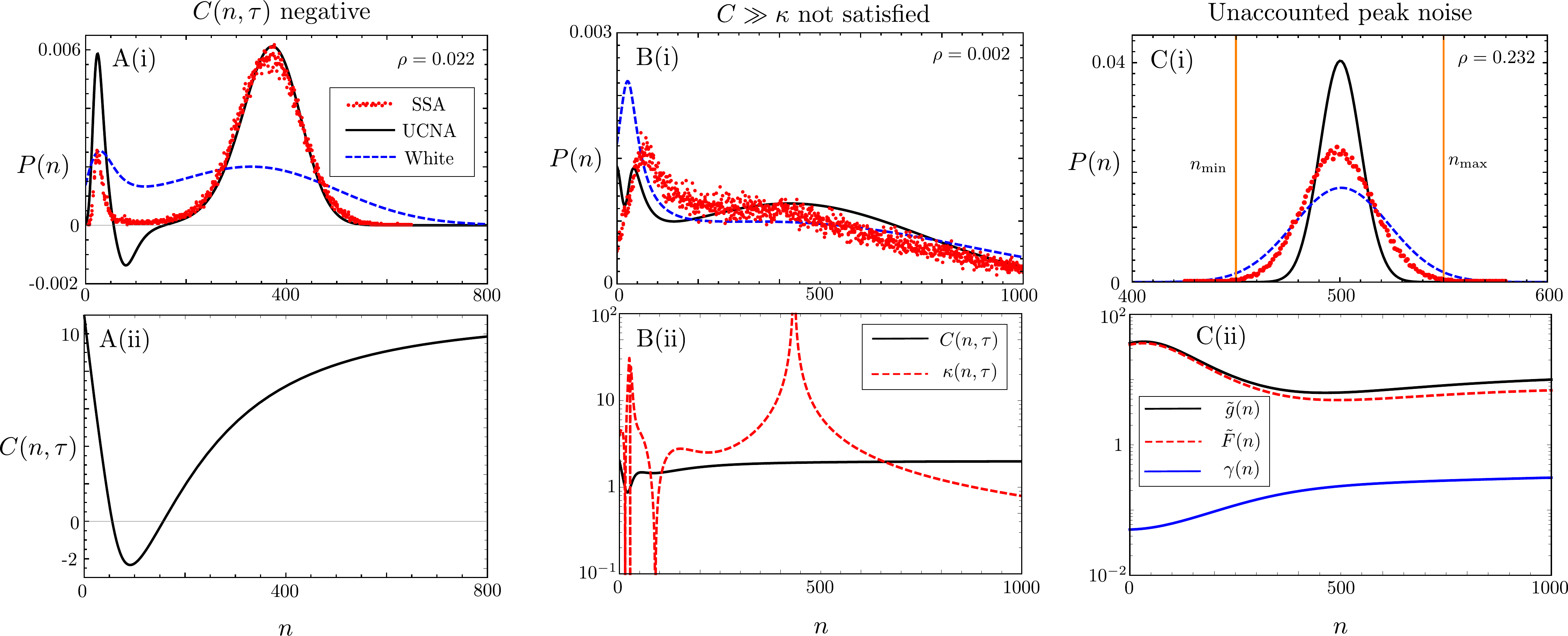}
\caption{This figure shows the disagreement of the UCNA on the protein production rates to the ground truth modified SSA predictions (detailed in Section \ref{sec:cnpp}), also compared to the case of white extrinsic noise ($\tau=0$). Each disagreement corresponds to a single breakdown condition of the UCNA being violated. The legend in A(i) applies to A(i), B(i) and C(i). Plots in A show the breakdown of the UCNA due to condition 1. A(i) shows the prediction of negative probability due to the negativity of $C(n,\tau)$ in A(ii) around the same value of $n$. Parameters for A are $r_u^{(0)}=20$, $r_b^{(0)}=470$, $s_u=s_b=10^3$, $d=1$, $D_1=1$, $D_2=0.1$, $\tau=10$, $\Omega=200$. Plots in B show the breakdown of the UCNA due to condition 2. B(ii) shows that $\kappa(n,\tau)>C(n,\tau)$ over a large range of $n$, corresponding to the poor UCNA prediction seen in B(i) over this entire region. Parameters for B are $r_u^{(0)}=50$, $r_b^{(0)}=450$, $s_u=s_b=10^3$, $d=1$, $D_1=1$, $D_2=1$, $\tau=1$, $\Omega=100$. Plots in C show the breakdown of the UCNA due to condition 3. C(ii) shows a relatively large value of $\gamma(n)$ over most of the defined region $\mathcal{D}$, and also shows the the pre-factors of the total UCNA noise $\Tilde{g}(n)$ and that arising only from the colored noise $\Tilde{F}(n)=F(n)/C(n,\tau)$. Vertical orange lines in C(i) indicate the limits of the region $\mathcal{D}$. The value $\rho$ in the top right-hand corner of C(i) can be compared to the smaller values of $\rho$ seen for other parameter sets in A(i) and B(i), indicating that the breakdown observed is truly associated to condition 3. The parameters for C are $r_u^{(0)}=2300$, $r_b^{(0)}=120$, $s_u=s_b=10^4$, $d=1$, $D_1=0.002$, $D_2=0.04$, $\tau=2$, $\Omega=230$. SSA data in each case comes from a single steady state trajectory of $9\times 10^5$s.}
\label{fig7}
\end{figure}

The first of these conditions concerns the positivity condition required on $C(n,\tau)$, that is $C(n,\tau)>0\,\forall \,n$. We refer to this as \textit{condition 1}. Since we have already discussed this condition in a previous section we will not repeat the discussion here, and refer the reader to Section \ref{sec:fdr}. In Fig. \ref{fig7}A(i) we see a disagreement between the UCNA and the modified SSA for a parameter set that exhibits bimodality, and in Fig. \ref{fig7}A(ii) it is verified that this since $C(n,\tau)<0$ where $n \approx 100$. Note however, that if $C(n,\tau)$ becomes negative outside of the region containing most of the probability mass that the UCNA can still provide a good approximation to the true modified SSA solution.

\subsubsection{Condition 2}

The second condition observed for the breakdown of the UCNA concerns the violation of the \textit{characteristic `length' scale} (the length here being a distance measure in the $n$ space), which we now discuss. In Appendix \ref{cond2} we show in more detail why the arguments we present below hold. Based on the noise intensity of the noise term arising from the colored noise in Eq. (\ref{eq:eppPreLE}), we can introduce the characteristic length scale $L$ over which fluctuations in the colored noise term are damped:
\begin{align}
    L(n,\tau) = \frac{F(n)}{C(n,\tau)},
\end{align}
noting that the requirement of condition 1 means that this length is always positive. Our approximate one variable FPE in Eq. (\ref{eq:prodFP}) will then be valid under the condition that the drift term varies slowly with respect to $L$ (following Appendix \ref{cond2}), meaning that one needs to satisfy
\begin{align}
    L\left|\partial_n \left(\Tilde{h}(n)+\Tilde{g}(n)\Tilde{g}'(n)\right)\right|\ll \left|\Tilde{h}(n)+\Tilde{g}(n)\Tilde{g}'(n)\right|
\end{align}
in order for the UCNA to hold. More succinctly, this condition is: 
\begin{align}\label{cond2eq}
    C(n,\tau)\gg \kappa(n,\tau),
\end{align}
where we henceforth define the function 
\begin{align}
\kappa(n,\tau) = F(n)\left|\frac{\partial_n \left(\Tilde{h}(n)+\Tilde{g}(n)\Tilde{g}'(n)\right)}{\Tilde{h}(n)+\Tilde{g}(n)\Tilde{g}'(n)}\right|.
\end{align}
We refer to Eq. (\ref{cond2eq}) as \textit{condition 2}. In Fig. \ref{fig7}B we explore this breakdown for a parameter set that breaks condition 2 over a large region of the parameter space, between $0<n<650$. Clearly the UCNA is provides a poor approximation in this regime; note however that, similar to condition 1, if condition 2 is violated (i) outside of the domain where most of the probability mass is contained, or (ii) over a small region of the domain containing most of the probability mass, then the UCNA can still provide a good approximation.

\subsubsection{Condition 3}

The final condition resulting in the breakdown of the UCNA concerns the underestimation of noise. We refer to this as \textit{unaccounted peak noise}, and this forms our final breakdown condition, \textit{condition 3}. The explanation behind condition 3 is that the UCNA in general will always underestimate the Poisson noise for a particular value of $n$, arising from the necessary neglection of Poisson noise terms in the derivation of the UCNA: (i) neglection of the noise terms under the square root of the Poisson noise pre-factor in Eqs. (\ref{eq:colLE2}) (a form of mean-field approximation), and (ii) neglection of Poisson noise term $g_2(n)\Gamma(t)$ and its time derivative from the $\dot{\eta}$ term in Eqs. (\ref{eq:etadeg}--\ref{mfdot_A}) via the use of another mean-field approximation. However, the error on the UCNA caused by condition 3 will be small when colored noise dominates the Poisson noise. To investigate the degree to which colored noise is dominant, identifying $F(n)/C(n,\tau)$ from Eq. (\ref{midwaySDE}) and $g(n)/C|(n,\tau)$ from Eq. (\ref{eq:UCNAtransLE}), we define
\begin{align}
    \gamma(n)=\left|\frac{g(n)/C(n,\tau) -F(n)/C(n,\tau)}{g(n)/C(n,\tau)} \right|=\left|\frac{g(n)-F(n)}{g(n)} \right|
\end{align}
where, for some $n$, if $\gamma(n)\approx 1$ then Poisson noise dominates, else if $\gamma(n)\approx 0$ then colored noise dominates. Intermediate values of $\gamma(n)$ mean that both Poisson and colored noise is apparent in the system. To investigate whether noise is underestimated generally over the region containing most of the probability, defined as $\mathcal{D}=[n_{\text{min}},n_{\text{max}}]$, we further define
\begin{align}
    \rho = \frac{1}{|\mathcal{D}|}\int_{n_{\text{min}}}^{n_{\text{max}}}\gamma(n) dn.
\end{align}
Here, if $\rho\approx 1$ then Poisson noise dominates over the entire region $\mathcal{D}$, else if $\rho\approx 0$ then colored noise dominates over the entire region $\mathcal{D}$. Fig. \ref{fig7}C explores this disagreement, where Fig. \ref{fig7}C(i) shows the clear underestimation of noise in the UCNA distribution when compared to the modified SSA distribution. Sample values of $n_{\text{min}}$ and $n_{\text{max}}$ are also shown on Fig. \ref{fig7}C(i). Fig. \ref{fig7}C(ii) shows how the total UCNA noise $\Tilde{g}(n)$ varies with respect to the contribution of colored noise $\Tilde{F}(n)=F(n)/C(n,\tau)$. Also shown on Fig. \ref{fig7}C(ii) is the variation of $\gamma(n)$. Values of $\rho$ are shown in the top right-hand corner for all probability distributions in Fig. \ref{fig5}; unlike the other distributions shown in Fig. \ref{fig7}, in Fig. \ref{fig7}C(i) $\rho\not\approx 0$ does not hold, clarifying that the reason for the UCNA's disagreement for this parameter set is due to condition 3.

\subsubsection{Large $\tau$ UCNA distributions}

\begin{figure}[h!]
\centering
\includegraphics[width=1.0\textwidth]{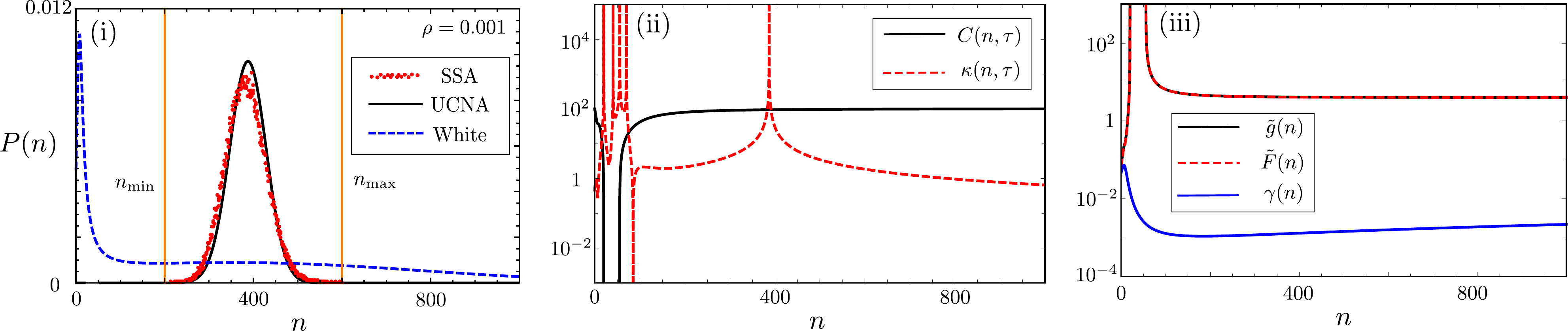}
\caption{Plots showing a good performance of the UCNA for $\tau=100$. (i) Shows the probability distributions from the modified SSA, UCNA and white noise approximation. Ther vertical orange lines show the limits of the region $\mathcal{D}$ in this case. (ii) Shows that in the region $\mathcal{D}$ that condition 1 is satisfied since $C(n\in \mathcal{D})>1$, and condition 2 is satisfied since $C(n\in \mathcal{D},\tau)\gg\kappa(n\in \mathcal{D},\tau)$. (iii) Shows that condition 3 is satisifed in $\mathcal{D}$, i.e., $\gamma(n)\approx 0$, since $\Tilde{g}(n)\approx \Tilde{F}(n)$, which is corroborated by the small value of $\rho$ shown in (i). Other parameters here are $r_u^{(0)}=10$, $r_b^{(0)}=400$, $s_u=s_b=10^3$, $d=1$, $D_1=0.5$, $D_2=1$, $\Omega=70$. SSA data in each case comes from a single steady state trajectory of $9\times 10^5$s.}
\label{fig8}
\end{figure}

Having successfully identified the three main conditions causing the breakdown of the UCNA, we are now able to determine where the UCNA will perform well, \textit{even in the large $\tau$ limit}. In Figure \ref{fig8} we explore an example of the UCNA performing exceptionally well for $\tau=10^2$ (see Fig. \ref{fig8}(i)). Clearly the UCNA does not violate any of the three conditions here: (1) $C(n,\tau)$ is not negative in $\mathcal{D}$ (see Fig. \ref{fig8}(ii)); (2) $C(n,\tau)\gg \kappa(n,\tau)$ in $\mathcal{D}$ (again, see Fig. \ref{fig8}(ii)); (3) $\gamma(n)$ is small for all $n$ in $\mathcal{D}$ (see Fig. \ref{fig8}(iii)) as evidenced by the small value of $\rho = 0.001$. As expected, the prediction of white noise on the protein production rates is very poor in the regime where $\tau\gg 1$.

\section{Slow gene switching: the conditional UCNA}\label{cUCNA}
In the previous sections we have focused on fast gene switching, whereby a Hill function can then be used to approximate the production of proteins from two different gene states, shown in the reaction scheme of Eq. (\ref{eq:coopRS}). We now consider the case where the switching rates $s_u$ and $s_b$ are very small; small enough that the system has two dominant modes of behaviour, one pertaining to each gene state. The approach followed here is very similar to the \textit{conditional linear noise approximation} (cLNA) studied in \cite{thomas2014phenotypic}, but instead of approximating the distribution conditional on each gene state as a Gaussian we instead utilise the UCNA in each gene state. We shall refer to this method as the \textit{conditional UCNA} (cUCNA). We begin by stating Bayes' theorem for the marginal distribution of proteins that we are interested in approximating:
\begin{align}\label{bayes}
    P(n,t) = \sum_{\underline{G}}P(G_i,t)P(n|G_i,t).
\end{align}
Here $\underline{G}$ is the set of possible gene state (in our case $\underline{G}=\{G,G^*\}$), $P(G_i,t)$ is the marginal distribution of being in gene state $G_i$ at a time $t$ and $P(n|G_i,t)$ is the conditional probability of having $n$ proteins at a time $t$ given that the system is in state $G_i$. Our task now is to find suitable approximations for $P(G_i,t)$ and $P(n|G_i,t)$ that allow us then to construct an approximation of the full steady state distribution in Eq. (\ref{bayes}). In our case we have two different gene states, $G$ and $G^*$, and hence we can construct the reaction schemes conditional on each gene state. The reaction scheme conditional on gene state $G$ is (i) $G \xrightarrow[]{r_u}G + P,\, P \xrightarrow{d} \varnothing$, and the reaction scheme conditional on gene state $G^*$ is (ii) $G^* \xrightarrow[]{r_b}G^* + P,\, P \xrightarrow{d} \varnothing$. This then allows us to approximately find the steady state mean number of proteins conditional on each gene state when $s_u$ and $s_b$ are very small (where the subscript $a$ denotes approximate): $\langle n|G\rangle_a = r_u/d$ and $\langle n|G^*\rangle_a = r_b/d$. We can use these conditional means to find the marginal probabilities of being in a specific gene state at steady state. Note that in this calculation we will ignore the influence of noise on the rate parameters; the inherent assumption is that extrinsic noise does not much influence the probability of being in each gene state. First we write an approximate master equation for the transitions between differing gene states:
\begin{align}
    \frac{d}{d t}P(G,t) \approx s_u P(G^*,t) - \frac{s_b \langle n|G\rangle_a^2}{\Omega^2}P(G,t).
\end{align}
We can then solve the above equation at steady state (denoted by the subscript $s$) by utilising conservation of probability, $P_s(G) = 1-P_s(G^*)$, giving:
\begin{align}\label{condG}
    P_s(G^*) &=\left(1+\frac{s_u}{s_b}\left(\frac{d\cdot\Omega}{r_u}\right)^2\right)^{-1},\\\label{condG2}
    P_s(G) &= \left(1+\frac{s_b}{s_u}\left(\frac{r_u}{d\cdot\Omega}\right)^2\right)^{-1}.
\end{align}
Since now we have the $P_s(G_i)$ needed for Eq. (\ref{bayes}) we need to find the $P_s(n|G_i)$ terms. Here we show how to calculate these terms for noise on the degradation rate, although this can be easily extended to the case where we have noise on the protein production rates. In each gene state, the system we are concerned to study is $G_i \xrightarrow[]{r_i}G_i + P,\, P \xrightarrow{d} \varnothing$, where $G_i$ and $r_i$ represent either gene state $G$ or $G^*$ and the corresponding production rate $r_u$ or $r_b$ respectively. Adding colored noise to the degradation rate, $d=d_0(1+\eta)$, we then have the following set of SDEs in each gene state (here we have applied the mean field approximation to the terms in the square root):
\begin{align}
    \frac{dn}{dt} &= r_i-d_0\, n -(d_0\, n)\eta+\sqrt{r_i+d_0 n}\cdot\Gamma(t),\\
    \frac{d\eta}{dt} &= -\frac{1}{\tau}\eta + \frac{1}{\tau}\theta(t),
\end{align}
where $\Gamma(t)$ and $\theta(t)$ are Gaussian white noise terms, each with zero mean and correlators $\langle \Gamma(t) \Gamma(t') \rangle = \delta(t-t')$ and $\langle \theta(t) \theta(t') \rangle = 2 D \delta(t-t')$ respectively. Processing the usual steps of the UCNA method, detailed explicitly in Section \ref{sec3}, we find the approximate steady state probability for each gene state:
\begin{align}\label{condN}
    P_s(n|G_i)\approx N\exp\left(u(n,r_i)\right)n^{2 r_i \tau-1}(r_i+d_0\,n(2 d_0 D n+1))^{-\frac{1}{2}-\frac{1}{2 d_0 D}- r_i \tau}(n+r_i \tau),
\end{align}
where $N$ is a normalisation constant and we have defined,
\begin{align}\label{udef}
    u(n,r_i) = \frac{\left(D \left(4-6 d_0 \tau \right) r_i+1\right) \tan ^{-1}\left(\frac{4 d_0 D n+1}{\sqrt{8 D r_i-1}}\right)}{d_0 D \sqrt{8
   D r_i-1}}.
\end{align}
We note that since each gene state is individually considered, each can have it's own associated noise strength $D$ and correlation time $\tau$ on the degradation rate. Hence, using Eqs. (\ref{condG})-(\ref{condG2}) and (\ref{condN}) we can now approximate Eq. (\ref{bayes}) as:
\begin{align}
    P(n) \approx P_s(G) P_s(n|G) + P_s(G^*) P_s(n|G^*).
\end{align}

\begin{figure}[h!]
\centering
\includegraphics[width=1.0\textwidth]{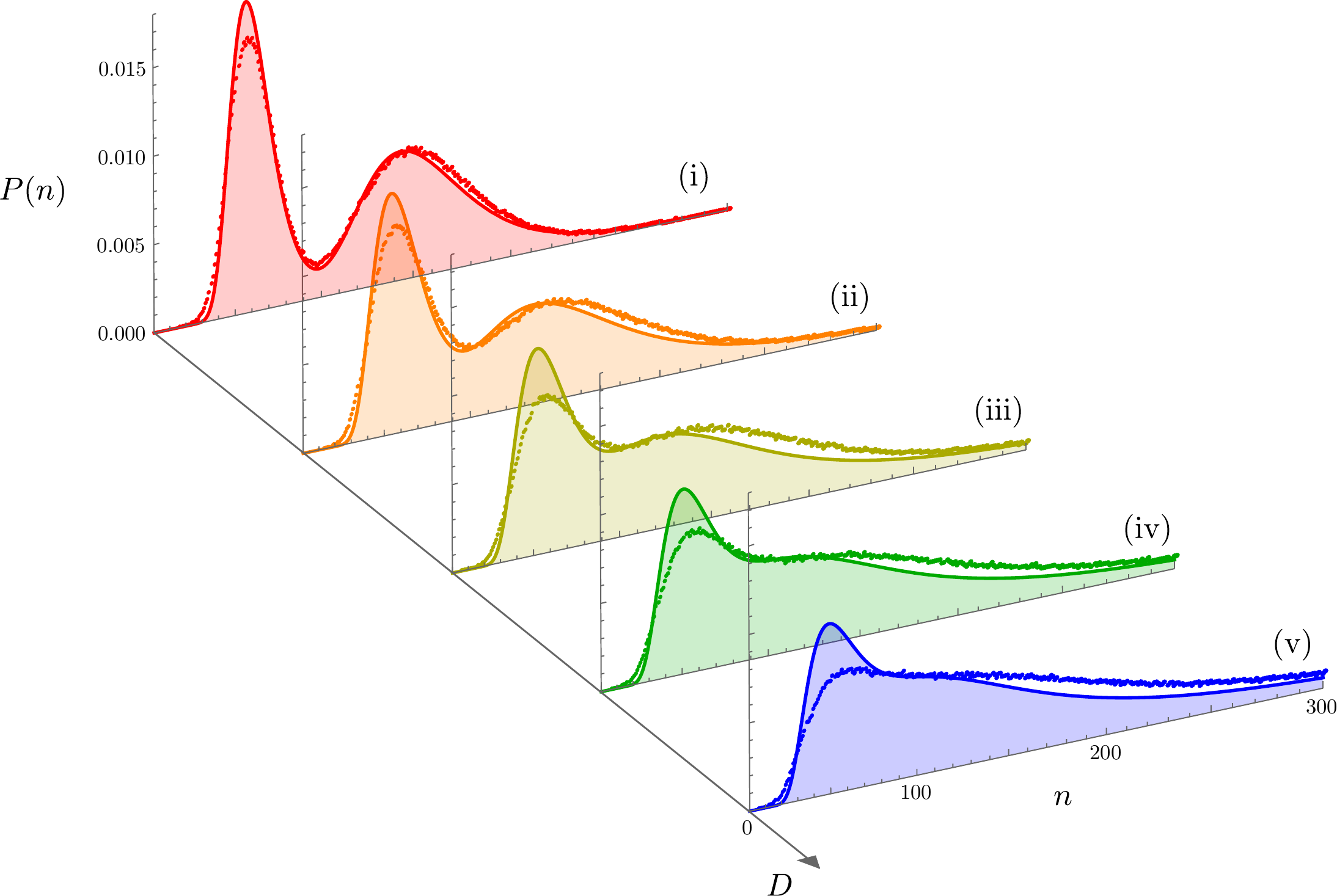}
\caption{Comparison of the cUCNA with the modified SSA for increasing values of the colored noise strength, $D$. It is seen that the cUCNA (solid lines) is a good approximation to the true distribution (dots, simulated using the modified SSA described in Appendix \ref{sec:ssa}), especially for small values of $D$. The noise strengths for each plot are (i) $D=0.1$, (ii) $D=0.2$, (iii) $D=0.3$, (iv) $D=0.4$, (v) $D=0.5$, and the shared parameters are $r_u=30$, $r_b=75$, $s_b=0.01$, $s_u=0.001$, $d=0.5$ and $\tau = 1$. Clearly as $D$ gets larger the bimodality exhibited by the slow switching between the gene states is destroyed by the extrinsic noise added to the degradation rate.}
\label{figCUCNA}
\end{figure} 

Figure \ref{figCUCNA} compares the cUCNA with the modified SSA -- which is the same as the modified SSA found in Section \ref{sec:fdr}. Fig. \ref{figCUCNA}(i) shows that for small switching rates, the cUCNA can correctly capture the bimodality exhibited where the colored noise on the degradation rate is small. As the noise on the degradation rate gets larger the cUCNA still provides a decent approximation to the true distributions; it is also clear that the bimodality of the protein distribution is destroyed as the size of this noise increases. One can contrast this to the cases observed in Figs. \ref{fig1} and \ref{fig6} which showed that where the gene switching rates are fast, increased colored noise strength can in fact induce bimodality. {\it{In summary, we find that extrinsic noise on the degradation rate of a slow switching auto-regulatory system generally destroys bimodality, but for fast switching it is common to observe the opposite phenomenon.}}

\section{Applications}\label{sec:app}
In this section we explicitly show, by means of two examples, how one can use the colored noise formulation that was introduced earlier to describe intricate molecular details of cooperative autoregulation. We first show this for multi-stage protein production with fast gene switching, and then for multi-stage protein degradation with slow gene switching. 

\subsection{Multi-stage protein production}

The first example of using colored noise as a form of model reduction is that of mapping multistage protein production onto a simpler system, where colored noise accounts for processes not explicitly considered in the simpler model. Consider multi-stage protein production on the cooperative auto-regulatory feedback loop:
\begin{align}\label{fullsys}
    &G\xrightarrow{\rho_u}G+M_1,\,G^*\xrightarrow{\rho_b}G^*+M_1,\\\nonumber &M_i\xrightarrow{\Lambda_i}M_{i+1}, \, i \in [1,N-1],\, M_N\xrightarrow{\Lambda_N}\varnothing,\, M_N\xrightarrow{r_1}M_N+P,\\\nonumber
    &G+2P\xrightleftharpoons[s_u]{s_b}G^*,\,P\xrightarrow{d}\varnothing,
\end{align}
where it is assumed the system contains only one gene, either in state $G$ or in state $G^*$. The simpler model that we will then map this system onto the cooperative auto-regulatory feedback loop:
\begin{align}\label{simplersys}
    &G \xrightarrow{r_u}G+P,\, G^* \xrightarrow{r_b}G^*+P,\\\nonumber & G+2P\xrightleftharpoons[s_u]{s_b}G^*,\,P\xrightarrow{d}\varnothing,
\end{align}
where $r_u=r_u^{(0)}(1+\eta_1(t))$ and $r_b=r_b^{(0)}(1+\eta_2(t))$, and assigning the properties of extrinsic noises $\eta_1(t)$ and $\eta_2(t)$ such that Eq. (\ref{fullsys}) can be mapped onto Eq. (\ref{simplersys}) is the task we have assigned ourselves. One can think of the different $M_i$ for $i<N$ as the various stages of nascent mRNA, before it is eventually fully transcribed in stage $M_N$ (mature mRNA) where it can then begin translation \cite{choubey2015deciphering}. Utilising the slow scale linear noise approximation \cite{thomas2012slow} one can show that if $\Lambda_i \gg \text{max}\{\Lambda_N,\rho_u,\rho_b\}$ for  $i\in[1,N-1]$ then the nascent mRNA $M_1,...,M_{N-1}$ are fast species, and the reaction system in Eq. (\ref{fullsys}) is consistent with the following reaction scheme describing fluctuations in the slow species $G$, $G^*$, $M_N$ and $P$:
\begin{align}\label{redscheme}
    &G\xrightarrow{\rho_u}G+M_N,\,G^*\xrightarrow{\rho_b}G^*+M_N,\\\nonumber &M_N\xrightarrow{\Lambda_N}\varnothing,\, M_N\xrightarrow{r_1}M_N+P,\\\nonumber
    &G+2P\xrightleftharpoons[s_u]{s_b}G^*,\,P\xrightarrow{d}\varnothing.
\end{align}
We now apply the van Kampen ansatz to the number of mature mRNA, $M_N$. In gene state $G$ this gives us $n_1(t) = \Omega\phi_1+\Omega^{1/2}\epsilon_1(t)$, and in gene state $G^*$ this gives us $n_2(t) = \Omega\phi_2+\Omega^{1/2}\epsilon_2(t)$, where $\phi_1=\rho_u/(\Lambda_N \Omega)$ and $\phi_2=\rho_b/(\Lambda_N \Omega)$ are the steady state solutions to the rate equation describing the mature mRNA in the gene states $G$ and $G^*$ respectively, and $\epsilon_1(t)$ and $\epsilon_2(t)$ describe small fluctuations about these means. Note the occurrence of $1/\Omega$ in $\phi_1$ and $\phi_2$ follows since the concentration of a single gene in a volume $\Omega$ is $1/\Omega$. Using these ansatzes allows us to construct the effective protein production rates in gene states $G$ and $G^*$ respectively:
\begin{align}
    r_u &= r_1 n_1(t)= \frac{r_1 \rho_u}{\Lambda_N}\left(1+\Omega^{1/2}\frac{\Lambda_N}{ \rho_u}\epsilon_1(t)\right),\\
    r_b &= r_1 n_2(t)= \frac{r_1 \rho_b}{\Lambda_N}\left(1+\Omega^{1/2}\frac{\Lambda_N}{ \rho_b}\epsilon_2(t)\right).
\end{align}
One can then see that $r_u^{(0)} = r_1 \rho_u/\Lambda_N$, $r_b^{(0)} = r_1 \rho_b/\Lambda_N$ and that the noise terms have the form:
\begin{align}\label{rel_eta_ep}
    \eta_1(t) &= \Omega^{1/2}\frac{\Lambda_N}{ \rho_u}\epsilon_1(t),\\\label{rel_eta_ep2}
    \eta_2(t) &= \Omega^{1/2}\frac{\Lambda_N}{ \rho_b}\epsilon_2(t).
\end{align}
In order to fully specify $\eta_1(t)$ and $\eta_2(t)$ we need to find the correlators $\langle \eta_1(t)\eta_1(t')\rangle$ and $\langle \eta_2(t)\eta_2(t')\rangle$, which can be done by application of the linear noise approximation (LNA) \cite{van1992stochastic}. Applying the LNA to $n_1(t)$ and $n_2(t)$, whose fluctuations are fully specified by the reactions $G\xrightarrow{\rho_u}G+M_N$, $G^*\xrightarrow{\rho_b}G^*+M_N$ and $M_N\xrightarrow{\Lambda_N}\varnothing$, gives us the two following one variable FPEs:
\begin{align}
    \frac{\partial \Pi(\epsilon_1,t)}{\partial t} &= \Lambda_N\frac{\partial}{\partial \epsilon_1}\left(\epsilon_1 \Pi(\epsilon_1,t)\right)+\frac{1}{2}\left(\frac{2 \rho_u}{\Omega}\right)\frac{\partial^2  \Pi(\epsilon_1,t)}{\partial \epsilon_1^2},\\
    \frac{\partial \Pi(\epsilon_2,t)}{\partial t} &= \Lambda_N\frac{\partial}{\partial \epsilon_2}\left(\epsilon_2 \Pi(\epsilon_2,t)\right)+\frac{1}{2}\left(\frac{2 \rho_b}{\Omega}\right)\frac{\partial^2  \Pi(\epsilon_2,t)}{\partial \epsilon_2^2},
\end{align}
where $\Pi(\epsilon_i,t)$ is the probability of having a fluctuation of size $\epsilon_i$ at a time $t$. These FPEs, combined with Eq. (\ref{rel_eta_ep}) and (\ref{rel_eta_ep2}), admit equivalent Langevin equations for $\eta_1(t)$ and $\eta_2(t)$, given by:
\begin{align}
    \frac{d\eta_1(t)}{dt} &= \Lambda_N \left(-\eta_1(t)+\sqrt{\frac{2}{\rho_u}}\beta_1(t)\right),\\
    \frac{d\eta_2(t)}{dt} &= \Lambda_N \left(-\eta_2(t)+\sqrt{\frac{2}{\rho_b}}\beta_2(t)\right),
\end{align}
where $\beta_1(t)$ and $\beta_2(t)$ are independent Gaussian white noises with zero mean and correlator $\langle \beta_1(t)\beta_1(t')\rangle=\langle \beta_2(t)\beta_2(t')\rangle=\delta(t-t')$. From here one can find the correlators of $\eta_1(t)$ and $\eta_2(t)$:
\begin{align}
    \langle \eta_1(t)\eta_1(t')\rangle &= \frac{\Lambda_N}{\rho_u}\exp\left(-\Lambda_N|t-t'|\right),\\
    \langle \eta_2(t)\eta_2(t')\rangle &= \frac{\Lambda_N}{\rho_b}\exp\left(-\Lambda_N|t-t'|\right).
\end{align}
Comparing to the results of Section \ref{sec:cnpp} it is clear that $\eta_1(t)$ and $\eta_2(t)$ satisfy the definition of colored noise, with noise strengths $D_1=1/\rho_u$, $D_2 = 1/\rho_b$ and shared correlation time $\tau = 1/\Lambda_N$. This completes the mapping between the full complex system in Eq. (\ref{fullsys}) and our reduced process in Eq. (\ref{simplersys}). We can hence utilise our solution for the probability distribution with colored noise on the effective protein production rates in Eq. (\ref{CNprodProb}).

\begin{figure}[h!]
\centering
\includegraphics[width=1.0\textwidth]{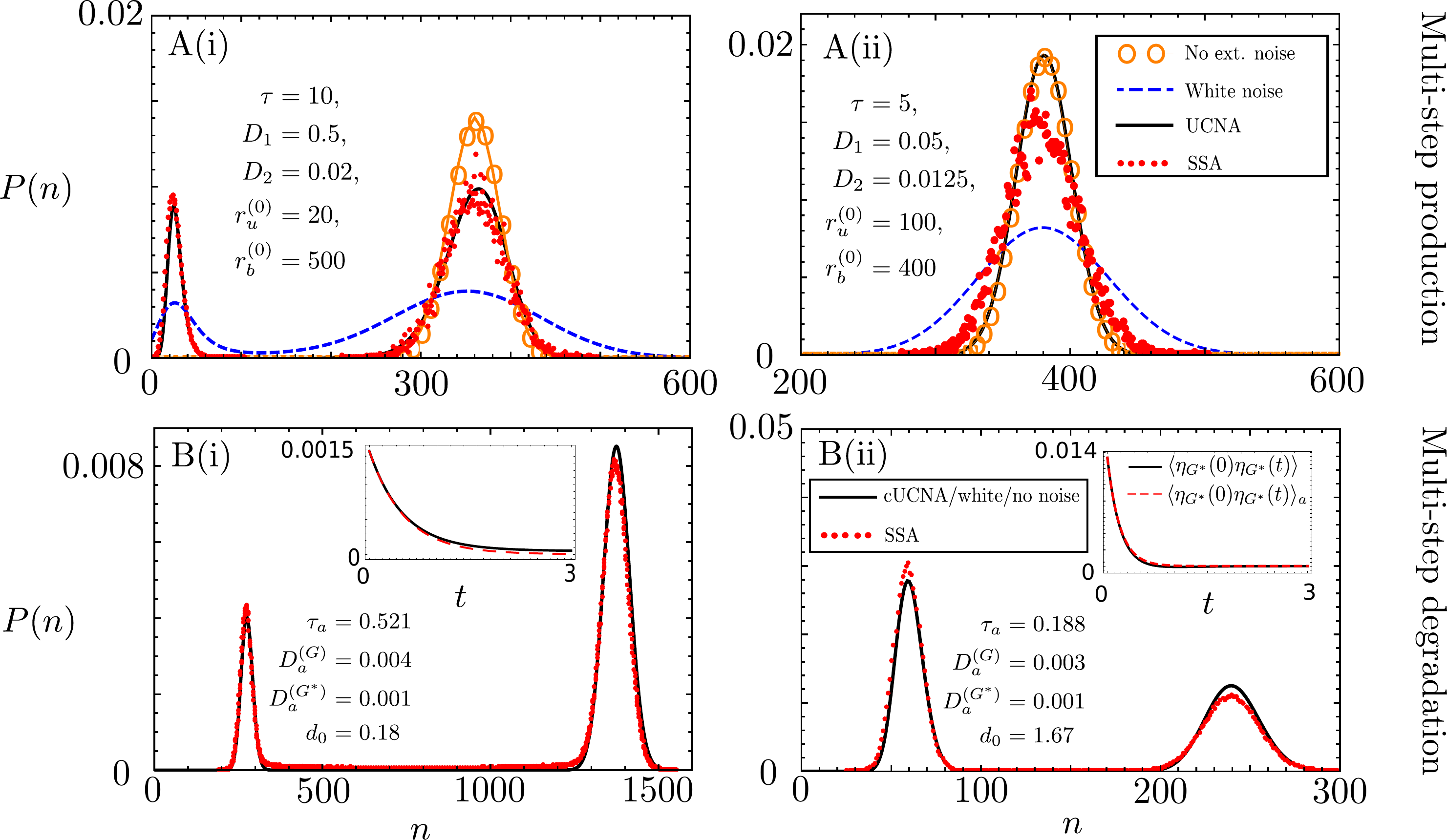}
\caption{(A) Shows distributions of the \textit{standard SSA} of the reaction scheme in Eq. (\ref{fullsys}) for multistage protein production in three intermediate species $M_1$, $M_2$ and $M_3$, against (1) the UCNA, (2) white extrinsic noise (i.e., $\tau=0$) and (3) no extrinsic noise. Each plot shows the colored noise parameters used for the UCNA solution from Eq. (\ref{CNprodProb}), which are determined from the full multi-stage protein production process in Eq. (\ref{fullsys}). Note the legend in A(ii) applies only to distributions in A(i--ii). Parameter values for the standard SSA in A(i) are $\rho_u = 2$, $\rho_b=50$, $\Lambda_1=1000$, $\Lambda_2=1000$, $\Lambda_3=0.1$, $r_1 = 1$, $s_u = s_b = 1000$, $d=1$ and $\Omega = 230$. Parameter values for the SSA in A(ii) are $\rho_u = 20$, $\rho_b=80$, $\Lambda_1=1000$, $\Lambda_2=1000$, $\Lambda_3=0.2$, $r_1 = 1$, $s_u = s_b = 1000$, $d=1$ and $\Omega = 100$. (B) Shows distributions of the \textit{standard SSA}, with protein decay following the multi-step process of Eq. (\ref{multistepDeg}), against the cUCNA. The colored noise parameters used for the cUCNA solution of Eq. (\ref{cucnaProbApp}) are shown on each plot, with these values being determined from the full model using Eqs. (\ref{effdegms},\ref{effDandTau1},\ref{effDandTau2}). The insets show a comparison of the approximate correlator (see Eq. (\ref{2expApp})) and the full double exponential correlator (see Eq. (\ref{2exp})) for gene state $G^*$. Note the legend in B(ii) applies only to distributions in B(i--ii), and the legend on the inset of B(ii) applies also to the inset in B(i). Parameter values for the SSA in B(i) are $r_u = 50$, $r_b=250$, $s_b = 2.5\times10^{-3}$, $s_u = 10^{-3}$, $\Omega=200$, $d_1 = 1$, $k=1$ and $d_2=0.1$. Parameter values for the SSA in B(ii) are $r_u = 100$, $r_b=400$, $s_b = 10^{-3}$, $s_u = 10^{-4}$, $\Omega=200$, $d_1 = 1$, $k=1$ and $d_2=5$. SSA data for A(i) and A(ii) come from a single steady state trajectories of length $10^8$s and $9\times 10^5$s respectively. Note that A(i) presents a very long relaxation to the steady state due to the systems inertia in staying in one of the two modes of the distribution. SSA data for B(i) and B(ii) come from a single steady state trajectory of $9\times 10^6$s.}
\label{figApp}
\end{figure}

In Fig. \ref{figApp}A we show how effective the UCNA can be in approximating the protein distribution from the full system described in Eq. (\ref{fullsys}), where we have for simplicity assumed that there are three mRNA states: $M_1$, $M_2$ and $M_3$ (i.e., $N=3$). Fig. \ref{figApp}A(i) shows the approximation for a parameter set exhibiting bimodality: the red points represent the \textit{standard SSA} of the full system in Eq. (\ref{fullsys}); the black line represents the distribution predicted from the UCNA (i.e., using Eq. (\ref{CNprodProb}) with $D_1=1/\rho_u$, $D_2=1/\rho_b$ and $\tau=1/\Lambda_N$); the blue dotted line represents the distribution if one put white noise of the same magnitude on the protein production rates (i.e., the UCNA at $\tau = 0$); and the orange line with circles shows the distribution if one was to neglect noise on the reaction rates entirely (i.e., $r_u = r_u^{(0)}$ and $r_b = r_b^{(0)}$). Clearly, in Fig. \ref{figApp}A(i) the UCNA is the only distribution that fits the SSA prediction, showing both the effectiveness of our model reduction as well as the need to properly account for the correlation time of colored noise in model reduction. This makes sense, since one would expect processes occurring in the full system to be correlated over short times, \textit{i.e., that noise events in close temporal proximity are not independent}, and one cannot simply neglect these effects. Fig. \ref{figApp}A(ii) instead shows the various approximations for a monomodal parameter set. In this case, white noise is a poor approximation, and it is clear that one cannot neglect the finite correlation time. However, it is interesting to note that properly accounting for the correlation time using the UCNA returns the same distribution as if one had not added noise to the production rates at all -- this is due to the small magnitudes of $D_1/\tau$ and $D_2/\tau$ respectively. These two examples shows that where correlation time is finite, it is imperative that one models it correctly.

\subsection{Multi-stage protein degradation}\label{sec:msdeg}

Proteins in cells are often degraded via multi-step processes. For example, a major degradation pathway in eukaryotic cells is the ubiquitin-proteasome degradation pathway \cite{cooper2000cell}, and more recent experiments have shown that a subset of proteins in the mammalian proteome have age-dependent degradation rates \cite{mcshane2016kinetic,pedraza2008effects}. From Fig. 2 in \cite{mcshane2016kinetic} we consider a system with two different stages of protein with differing degradation rates combined with the cooperative auto-regulatory feedback loop:
\begin{align}\label{multistepDeg}
    &G\xrightarrow{r_u}G+P_1,\, G^*_{1,2}\xrightarrow{r_b}G^*_{1,2}+P_1,\\\nonumber
    &P_1 \xrightarrow{\kappa} P_2,\, P_1 \xrightarrow{d_1} \varnothing, \, P_2 \xrightarrow{d_2}
    \varnothing,\\\nonumber
    & G+2P_1 \xrightleftharpoons[s_u]{s_b}G^*_{{1}},\, G+2P_2 \xrightleftharpoons[s_u]{s_b}G^*_{{2}},
\end{align}
where $G^*_{1,2}$ indicates \textit{either} the state $G^*_{1}$ or $G^*_{2}$. This reaction system models age dependent protein states, since the protein $P_1$ is always produced from the gene, and eventually undergoes a transition to protein state $P_2$, where $P_1$ and $P_2$ have differing degradation rates. We will show how to map this system to the reduced system:
\begin{align}\label{multistepDegSimple}
    G+2P \xrightleftharpoons[s_u]{s_b}G^*,\,G\xrightarrow{r_u}G+P,\, G^*\xrightarrow{r_u}G^*+P,\, P \xrightarrow{d}\varnothing,
\end{align}
where the total number of $P$ is given as the sum of the number of $P_1$ and $P_2$, i.e., $n=n_1+n_2$, $G^*$ is simply the sum of $G^*_{1}$ and $G^*_{2}$, and $d = d_0(1+\eta(t))$, where $\eta(t)$ is extrinsic noise. Our task is to find the properties of the noise $\eta(t)$ such that one can map the full system in Eq. (\ref{multistepDeg}) onto the reduced system in Eq. (\ref{multistepDegSimple}). Note that although we here look at two different stages of protein, the analysis presented below can be easily extended for several different stages of protein degradation. 

One finds that the effective degradation rate of the sum of protein number $P_1$ and $P_2$ is:
\begin{align}\label{effdeg}
    d = \frac{n_1 d_1+ n_2 d_2}{n_1+n_2}.
\end{align}
In the following analysis we consider gene switching to be slow, which allows us to apply the cUCNA from Section \ref{cUCNA}. We first consider the probability of being in each gene state at steady state $P_s(G_k)$, where $G_k$ represents either gene state $G$ or $G^*$. Note that we assume both protein stages $P_1$ and $P_2$ can bind and unbind to the gene at the same respective rates, and note that $\langle n|G\rangle = \langle n_1|G\rangle+\langle n_2|G\rangle$. Following the analysis from Section \ref{cUCNA} in Eqs. (\ref{condG}--\ref{condG2}) we find:
\begin{align}\label{gsweights1}
    P_s(G^*) &=\left(1+\frac{s_u \Omega^2}{s_b}\left(\frac{d_2(\kappa+d_1)}{r_u(\kappa+d_2)}\right)^2\right)^{-1},\\\label{gsweights2}
    P_s(G) &= \left(1+\frac{s_b}{s_u \Omega^2}\left(\frac{r_u(\kappa+d_2)}{d_2(\kappa+d_1)}\right)^2\right)^{-1}.
\end{align}

We can now proceed to find the probability distribution conditional on each gene state $P_s(n_1,n_2)$. In gene state $G_k$ the conditional reaction system is:
\begin{align}\label{RSdeg}
    G_k \xrightarrow[]{r_k}G_k + P_1,\, P_1 \xrightarrow{d_1}\varnothing ,\, P_1 \xrightarrow{\kappa} P_2, \, P_2 \xrightarrow{d_2} \varnothing,
\end{align}
where the protein is always produced in stage $P_1$, and $r_k\equiv r_u$ in gene state $G$, and $r_k\equiv r_b$ in gene state $G^*$. Now we employ the van Kampen ansatz \cite{van1992stochastic} on $n_1$ and $n_2$ in gene state $G_k$, i.e. $n_1^{(k)}(t)=\Omega \phi^{*(k)}_1+\Omega^{1/2}\epsilon_1^{(k)}(t)$ and $n_2^{(k)}(t)=\Omega \phi^{*(k)}_2+\Omega^{1/2}\epsilon_2^{(k)}(t)$, where $\phi^{*(k)}_1$ and $\phi^{*(k)}_2$ are the deterministic steady state mean concentrations of $P_1$ and $P_2$ in gene state $G_k$ respectively, and $\epsilon_1^{(k)}(t)$ and $\epsilon_2^{(k)}(t)$ are fluctuations about these mean values. In the following we drop the superscript $(k)$ notation for aesthetic reasons, although one should keep in mind that the process below must be individually conducted on each gene state. The purpose of using the van Kampen ansatz can be seen upon its substitution into Eq. (\ref{effdeg}) which for a large system size, $\Omega$, gives:
\begin{align}
    d = \frac{d_1 \phi_1^*+d_2\phi_2^*}{\phi_1^*+\phi_2^*}\Bigg(1+\Omega^{-1/2}\Big(\epsilon_1(t)&\Big(\frac{d_1}{d_1 \phi_1^*+d_2\phi_2^*}-\frac{1}{\phi_1^*+\phi_2^*}\Big)\\\nonumber + &\epsilon_2(t)\Big(\frac{d_2}{d_1 \phi_1^*+d_2\phi_2^*}-\frac{1}{\phi_1^*+\phi_2^*}\Big)\Big)\Bigg) + \mathcal{O}(\Omega^{-1}).
\end{align}
By comparing to the effective degradation from the reduced model in gene state $G_k$, $d = d_0(1+\eta_k(t))$, one can see that to match the two models we must have 
\begin{align}\label{effdegms}
    d_0 = (d_1 \phi_1^*+d_2\phi_2^*)/(\phi_1^*+\phi_2^*),
\end{align}
and
\begin{align}\label{degeta}
    \eta_k(t) = \Omega^{-1/2}\left(\epsilon_1(t)y_1(d_1,d_2,\phi_1^*,\phi_2^*) + \epsilon_2(t)y_2(d_1,d_2,\phi_1^*,\phi_2^*)\right),
\end{align}
where we have defined the functions,
\begin{align}
    y_1(d_1,d_2,\phi_1^*,\phi_2^*)&=\frac{d_1}{d_1 \phi_1^*+d_2\phi_2^*}-\frac{1}{\phi_1^*+\phi_2^*},\\
    y_2(d_1,d_2,\phi_1^*,\phi_2^*)&=\frac{d_2}{d_1 \phi_1^*+d_2\phi_2^*}-\frac{1}{\phi_1^*+\phi_2^*}.
\end{align}
If the correlators $\langle \epsilon_1(0)\epsilon_1(t)\rangle$, $\langle \epsilon_2(0)\epsilon_2(t)\rangle$, $\langle \epsilon_1(0)\epsilon_2(t)\rangle$ and $\langle \epsilon_2(0)\epsilon_1(t)\rangle$ are known, then one can also find the correlator of $\eta_k(t)$, i.e., $\langle\eta_k(0)\eta_k(t)\rangle$, given by:
\begin{align}\label{etacorreq}
    \langle\eta_k(0)\eta_k(t)\rangle = \frac{1}{\Omega}\left(y_1^2 \langle \epsilon_1(0)\epsilon_1(t)\rangle + y_2^2 \langle \epsilon_2(0)\epsilon_2(t)\rangle + y_1 y_2(\langle \epsilon_1(0)\epsilon_2(t)\rangle+\langle \epsilon_2(0)\epsilon_1(t)\rangle) \right).
\end{align}
Note in Eq. (\ref{degeta}) that if $d_1=d_2=d$, then the magnitude of $\eta_k(t)$ is zero for all $t$ since the system $P_1 \xrightarrow{\kappa} P_2,\, P_1 \xrightarrow{d} \varnothing, \, P_2 \xrightarrow{d} \varnothing$ is equivalent to $P\xrightarrow{d}\varnothing$ where one is only interested in the total number of proteins.

To proceed in finding $\eta_k(t)$ in Eq. (\ref{degeta}), we first need to find the steady state concentrations $\phi^*_1$ and $\phi^*_2$ from the deterministic rate equations. These are,
\begin{align}\label{degdetRE1}
    \frac{d \phi_1}{dt} &= \frac{r_k}{\Omega}-(\kappa+d_1)\phi_1,\\
    \frac{d \phi_2}{dt} &= \kappa \phi_1 -d_2 \phi_2,
\end{align}
where again the $1/\Omega$ in Eq. (\ref{degdetRE1}) follows since the concentration of a single gene in a volume $\Omega$ is $1/\Omega$. Enforcing the steady state condition allows us to find $\phi^*_1$ and $\phi^*_2$,
\begin{align}\nonumber
    \phi^*_1 = \frac{r_k}{\Omega(\kappa+d_1)},\,
    \phi^*_2 = \frac{\kappa r_k}{d_2(\kappa+d_1)}.
\end{align}
Note that the linear dependence of $\phi_1^*$ and $\phi_2^*$ on $r_k$ means that the effective degradation rate $d_0$ from Eq. (\ref{effdegms}) is independent of the gene state. Assuming that both $P_1$ and $P_2$ are numerous, we now proceed to the LNA \cite{van1992stochastic,elf2003fast} of the system in Eq. (\ref{RSdeg}), which will allow us to find the correlators $\langle \epsilon_1(0)\epsilon_1(t)\rangle$, $\langle \epsilon_2(0)\epsilon_2(t)\rangle$, $\langle \epsilon_1(0)\epsilon_2(t)\rangle$ and $\langle \epsilon_2(0)\epsilon_1(t)\rangle$. Where $\textbf{S}$ is the stoichiometric matrix, $\underline{\phi}=(\phi_1,\phi_2)$ and $\underline{f}(\underline{\phi})$ is the macroscopic rate vector one can computationally find the required matrices needed to perform the LNA: (i) the steady state Jacobian matrix $A_{ij}=d(\textbf{S}\cdot \underline{f}(\underline{\phi}))_j/d\phi_i|_{\underline{\phi}=\underline{\phi}^*}$, and (ii) the steady state diffusion matrix $(\textbf{B}\cdot\textbf{B}^{T})_{ij} = \textbf{S}\cdot \text{Diag}(\underline{f}(\underline{\phi}))\cdot \textbf{S}^T|_{\underline{\phi}=\underline{\phi}^*}$. The Jacobian matrix then allows us to find the time evolution of both $\langle\epsilon_1(t)\rangle$ and $\langle\epsilon_2(t)\rangle$ since $\partial_t \langle\underline{\epsilon}\rangle = \textbf{A}\cdot\langle\underline{\epsilon}\rangle$, where $\langle\underline{\epsilon}\rangle=(\langle\epsilon_1(t)\rangle,\langle\epsilon_2(t))\rangle$. Solving these coupled first order ODEs gives us:
\begin{align}\label{ep1sol}
    \langle\epsilon_1(t)\rangle &= \langle\epsilon_1(0)\rangle e^{-(d_1+\kappa)t},\\
    \langle\epsilon_2(t)\rangle &= \frac{-\kappa \langle\epsilon_1(0)\rangle e^{-(d_1+\kappa) t}+(\kappa\langle\epsilon_1(0)\rangle+(d_1-d_2+\kappa)\langle\epsilon_2(0)\rangle)e^{-d_2 t}}{d_1-d_2+\kappa},
\end{align}
where $-d_2$ and $-(d_1+\kappa)$ are eigenvalues of $\textbf{A}$. Clearly, in the limit $t\to\infty$ the fluctuations about the steady state concentrations $\phi_1^*$ and $\phi_2^*$, $\langle\epsilon_1(t)\rangle$ and $\langle\epsilon_2(t)\rangle$, tend to zero as required. The final step of the LNA then requires us to find the covariance matrix $\textbf{C}$ at steady state, which has the steady state variances $\langle\epsilon_1^2\rangle$ and $\langle\epsilon_2^2\rangle$ as diagonal components and covariance $\langle\epsilon_1\epsilon_2\rangle=\langle\epsilon_2\epsilon_1\rangle$ in the off-diagonal components. $\textbf{C}$ is then given by the Lyapunov equation \cite{elf2003fast}:
\begin{align}
    \textbf{A}\cdot\textbf{C}+\textbf{C}\cdot\textbf{A}^T + \textbf{B}\cdot\textbf{B}^{T} = 0,
\end{align}
whose solution is given by:
\begin{align}\label{Csolmat}
    \textbf{C} = 
    \begin{pmatrix}
    \frac{r_k}{(d_1+\kappa)\Omega} & 0\\
    0 & \frac{\kappa r_k}{d_2(d_1+\kappa)\Omega}
    \end{pmatrix}.
\end{align}
From van Kampen \cite{van1992stochastic} p. 259 we assert that for some fluctuation $\epsilon_i$, $\langle \epsilon_i(0)\epsilon_j(t)\rangle = \langle \epsilon_i(0)\langle\epsilon_j(t)\rangle\rangle$, and that at $t=0$ we have $\underline{\phi} = \underline{\phi}^*$ so that $\langle\epsilon_i(0) \epsilon_j(0)\rangle = \langle \epsilon_i \epsilon_j\rangle$. For example, for $\langle \epsilon_1(0)\epsilon_1(t)\rangle$ we have, using $\langle \epsilon_1(t)\rangle$ from Eq. (\ref{ep1sol}) and $\langle \epsilon_1^2\rangle$ from Eq. (\ref{Csolmat}), $\langle \epsilon_1(0)\epsilon_1(t)\rangle = \langle \epsilon_1(0)\langle\epsilon_1(t)\rangle\rangle = \langle \epsilon_1^2\rangle e^{-(d_1+\kappa)t}$. Explicitly, one can then calculate all the correlators, which are given by:
\begin{align}
    \langle \epsilon_1(0)\epsilon_1(t)\rangle &= \frac{r_k}{(d_1+\kappa)\Omega}e^{-(d_1+\kappa)t},\\
    \langle \epsilon_2(0)\epsilon_2(t)\rangle &= \frac{\kappa r_k}{d_2(d_1+\kappa)\Omega}e^{-d_2 t},\\
    \langle \epsilon_2(0)\epsilon_1(t)\rangle &= 0,\\
    \langle \epsilon_1(0)\epsilon_2(t)\rangle &= \frac{\kappa r_k(e^{-d_2 t}-e^{-(d_1+\kappa)t})}{(d_1+\kappa)(d_1-d_2+\kappa)\Omega}.
\end{align}
Now that these correlators have been determined, we can substitute them into Eq. (\ref{etacorreq}) giving us the following for the correlator of $\eta_k(t)$:
\begin{align}\label{2exp}
    \langle \eta_k(0)\eta_k(t)\rangle = \frac{\left(d_1-d_2\right){}^2 \kappa \left(\kappa \left(d_1+\kappa\right) e^{-(d_1+\kappa) t}+\left(d_1-d_2\right) d_2 e^{-d_2 t}\right)}{\left(d_1+\kappa\right) \left(d_1-d_2+\kappa\right) \left(d_2+\kappa\right){}^2 r_k},
\end{align}
noting the only dependence on the gene state $G_k$ comes from the pre-factor $1/r_k$. Comparing this equation to the colored noise seen in Eq. (\ref{eq:colLE1}) in Section \ref{sec:fdr} we see however that we have two exponentials in the correlator. This sum of exponentials in Eq. (\ref{2exp}) can be approximated by a single exponential through a small $t$ expansion. This gives us:
\begin{align}\label{2expApp}
    \langle \eta_k(0)\eta_k(t)\rangle \approx \langle \eta_k(0)\eta_k(t)\rangle_a = \frac{D^{(k)}_a}{\tau_a}e^{-t/\tau_a},
\end{align}
where $D^{(k)}_a$ and $\tau_a$ are the approximate noise strength and correlation time given by:
\begin{align}\label{effDandTau1}
    D^{(k)}_a &= \frac{\left(d_1-d_2\right){}^2 \kappa}{\left(d_1+\kappa\right) \left(d_1 \kappa+d_2 \left(d_2+\kappa\right)+\kappa^2\right) r_k},\\\label{effDandTau2}
   \tau_a &= \frac{d_2+\kappa}{d_1 \kappa+d_2 \left(d_2+\kappa\right)+\kappa^2}.
\end{align}
Clearly, the small $t$ expansion allows us to roughly interpret the noise $\eta_k(t)$, present in gene state $G_k$, as colored noise with strength $D_a/\tau_a$ and correlation time $\tau_a$. Note that even when both exponentials equally contribute to the correlator in Eq. (\ref{2exp}), this is generally a very good approximation for few protein stages. Knowing $D^{(k)}_a$ and $\tau_a$ for $\eta_k(t)$ we can now substitute them into Eqs. (\ref{condN}--\ref{udef}) in Section \ref{cUCNA}, then using Eqs. (\ref{gsweights1}--\ref{gsweights2}) we find
\begin{align}\label{cucnaProbApp}
    P(n) \approx P_s(G) P_s(n|G) + P_s(G^*) P_s(n|G^*).
\end{align}

Fig. \ref{figApp}B shows two different cases of the cUCNA for predicting distributions for multi-stage degradation and slow gene switching: in B(i) for the case of $d_1>d_2$ (true for around 80\% of proteins in \cite{mcshane2016kinetic}); in B(ii) for the case of $d_2>d_1$ (true for around 20\% of proteins in \cite{mcshane2016kinetic}). On the main plots red dots show the standard SSA prediction of the full reaction scheme in Eq. (\ref{multistepDeg}), and the black lines show the cUCNA from Eq. (\ref{cucnaProbApp}), which in both cases is almost indistinguishable from the white noise (cUCNA with $\tau=0$) and no extrinsic noise prediction (discussed further in the following paragraph). The insets show the correlators in gene state $G^*$, where the red dashed line represents $\langle \eta_{G^*}(0)\eta_{G^*}(t)\rangle_a$ and the black line shows $\langle \eta_{G^*}(0)\eta_{G^*}(t)\rangle$. Note that the correlators for gene state $G$ are not shown because they show very similar to what is seen for state $G$. Even given the complex model reduction from two protein species to one effective protein species the cUCNA performs very well in predicting distributions from the standard SSA of the full system in Eq. (\ref{multistepDeg}). Note that as one considers more protein stages with differing degradation rates it becomes more different to fit the correlator to a single exponential, which presents a limitation of this method for more protein stages.

However, we find that since our analysis is restricted to the large protein number regime, and the noise strength $D_a$ is inversely proportional to the production rate $r_k$ which is typically large, that $D_a$ is typically very small in both gene states. This means that the cUCNA probability distribution is almost identical to probability distributions that assume white noise (cUCNA with $\tau=0$) or even no extrinsic noise. However, what the analysis in this section provides is the \textit{quantitative reasons why one could necessarily neglect the contribution of extrinsic noise in model reduction from the full system in Eq. (\ref{multistepDeg}) to the simpler system in Eq. (\ref{multistepDegSimple})}.

\section{Conclusion}\label{conc}

In this paper we have explored the addition of colored noise onto the reaction rates for a cooperative auto-regulatory circuit. Starting from a reduced chemical Fokker-Planck description, we used the UCNA to derive approximate expressions for the probability distribution of protein numbers in the limits of fast and slow promoter switching. The approximation is valid provided the colored noise on the reaction rates is small and the correlation time is short. By means of stochastic simulations, we verified the accuracy of the approximate distributions; we also verified the predictions of the UCNA, namely that under fast promoter switching conditions the addition of colored noise can induce bimodality whereas under slow promoter switching conditions, noise can destroy bimodality. 

We also have shown how complex models of gene expression can be mapped onto simpler models with noisy rates. In particular we have shown that: (i) An auto-regulatory feedback loop with multi-stage protein production, including different stages of mRNA processing, can be mapped onto an auto-regulatory feedback loop with a single protein production reaction step having colored noised added to its reaction rate. (ii) A feedback loop with multi-stage protein degradation can be mapped onto a feedback loop with a single protein degradation reaction with a fluctuating rate. We have also verified that in many instances, one cannot simply approximate colored noise with white noise, or else neglect it entirely, since this does not match behaviour seen from the full underlying models of multi-stage protein production or degradation.

The UCNA provides an easily extendable analysis for other gene regulatory networks so long as only one species effectively describes the system. Our analysis is the first to our knowledge, to analytically find steady state probability distributions where colored noise is added to a non-linear reaction (the protein-gene binding reaction) in a gene regulatory context; a previous study applied the UCNA to study the effects of extrinsic noise in genetic circuits composed of purely linear reactions \cite{shahrezaei2008colored}. Given that our calculations show that the protein distributions for auto-regulatory circuits with extrinsic noise on reaction parameters can be dramatically different than models assuming constant reaction rates, an interesting future research direction would be to develop UCNA based methods that can directly infer the properties of colored noise on reaction rates from protein expression data.    

\section*{Acknowledgments}{This work was supported by a BBSRC EASTBIO PhD studentship for J.H., a Darwin Trust scholarship for A.G. and a Leverhulme Trust grant (RPG-2018-423) for R.G.}

\appendix

\section{Stochastic simulations of autoregulation with extrinsic noise}\label{sec:ssa}
In this paper extrinsic noise is accounted for in the SSA through the introduction of a new ghost species $Y$ and some new ghost reactions. For example, consider the case where we want to model a fluctuating degradation rate $d=d_0(1+\eta(t))$, where $\langle \eta(t)\rangle = 0$ and $\langle \eta(t)\eta(t') \rangle = (D/\tau)\exp\left(-|t-t'|/\tau\right)$. We will then replace the degradation reaction, $P \xrightarrow[]{d}\varnothing$, by the following set of reactions:
\begin{align}\label{eq:SSA1}
    \varnothing \xrightleftharpoons[1/\tau]{1/(D\Omega)}Y,\;P+Y \xrightarrow[]{d_0 D \Omega/\tau} Y.    
\end{align}
We now show why this equivalence exists. The \textit{propensity} for the degradation reaction in Eq. (\ref{eq:SSA1}) is $(n_P n_Y d_0 D \Omega/\tau)/\Omega$, meaning that the \textit{effective degradation rate} is $d = (n_Y d_0 D \Omega/\tau)/\Omega$. Assuming there are large numbers of $Y$, we apply the van Kampen ansatz that fluctuations in $Y$ occur around its deterministic steady state mean \cite{van1992stochastic}:
\begin{align}\label{eq:ssaeq1}
    \frac{n_Y}{\Omega}&=\frac{\tau}{D\Omega}+\Omega^{-1/2}\epsilon(t).
\end{align}
Then, employing the system size expansion, and enforcing the linear noise approximation (LNA), we obtain a linear FPE for the probability of having a fluctuation of size $\epsilon(t)$ at a time $t$, denoted $\Pi(\epsilon,t)$ \cite{van1992stochastic,elf2003fast}:
\begin{align}
\frac{\partial\Pi(\epsilon,t)}{\partial t} = \frac{1}{\tau}\frac{\partial}{\partial \epsilon}(\epsilon \Pi(\epsilon,t))+\frac{1}{2}\frac{2}{D \Omega}\frac{\partial^2 \Pi(\epsilon,t)}{\partial \epsilon^2}.    
\end{align}
This FPE then admits an equivalent Langevin equation given by:
\begin{align}
    \frac{d \epsilon(t)}{d t} &= -\frac{1}{\tau}\epsilon(t)+\sqrt{\frac{2}{D \Omega}}\beta(t),
\end{align}
where $\beta(t)$ is Gaussian white noise with zero mean and correlator $\langle \beta(t)\beta(t') \rangle = \delta(t-t')$. Hence, from Eq. (\ref{eq:ssaeq1}) it follows that $d$ goes as:
\begin{align}\label{eq:de1}
    d &= d_0 D \frac{\Omega}{\tau}\frac{n_Y}{\Omega}=d_0(1+ \eta(t)),\\\label{eq:de2}
    \frac{d \eta(t)}{dt} &= -\frac{1}{\tau}\eta(t)+\frac{\sqrt{2 D}}{\tau}\beta(t),
\end{align}
where $\eta(t) = \Omega^{1/2} D \epsilon(t)/\tau$. Eqs. (\ref{eq:de1}) and (\ref{eq:de2}) are consistent with the definition of colored noise described at the beginning of this section. This modified SSA requires that where $\tau$ and $D$ are both individually large, that $\tau \gg D$ such that slow switching is not enforced between differing numbers of the ghost species.

\section{Detailed explanation of condition 2}\label{cond2}
In order to explain the origin of condition 2 -- a condition on the length scale of colored noise fluctuation compared to the rate of variation of the drift term in Eq. (\ref{eq:prodFP}) -- we will first consider a more intuitive example. Consider a Brownian particle subject to a force $F(x)$, whose state is specified by both its position $x$, as well as its velocity $v$. The set of SDEs governing the state of this particle is then \cite{van1992stochastic}:
\begin{align}
    \frac{dx}{dt} &= v,\\
    \frac{dv}{dt} &= \frac{F(x)}{m}-\gamma v + \frac{k_b T \gamma}{m}\Gamma(t),
\end{align}
where $m$ is the mass of the particle, $\gamma$ is the damping coefficient of the frictional force surrounding the particle (frictional force is $-\gamma m v$), $k_b T$ is the thermal energy of the particle, and $\Gamma(t)$ is Gaussian white noise with zero mean and correlator $\langle\Gamma(t)\Gamma(t')\rangle = \delta(t-t')$. The equivalent multivariate FPE for this set of SDEs is \cite{gardiner2009stochastic}:
\begin{align}\label{BMxv}
    \frac{\partial P(x,v;t)}{\partial t} = \gamma\left[\frac{\partial(v P)}{\partial v} + \frac{k_b T}{m}\frac{\partial^2 P}{\partial v^2} \right]-v\frac{\partial P}{\partial x}-\frac{F(x)}{m}\frac{\partial P}{\partial v}.
\end{align}
Now following van Kampen p. 216--218 \cite{van1992stochastic}, one can utilise singular perturbation theory assuming that the damping coefficient $\gamma$ is small (although the same procedure could be done for $\gamma$ large) in order to reduce the above FPE in two variables to a FPE in the position variable $x$ alone. The result after having done this procedure is:
\begin{align}\label{BMx}
    \frac{\partial P(x;t)}{\partial t} = -\frac{\partial}{\partial x}\left(\frac{F(x)}{m \gamma}P\right) + \frac{k_b T}{m \gamma}\frac{\partial^2 P}{\partial x^2}.
\end{align}
Aside from the requirement that $\gamma$ must be small, there is another condition required of Eq. (\ref{BMx}) such that it reasonably approximates Eq. (\ref{BMxv}). This condition arises \textit{physically} since we realise that if we are to approximate Eq. (\ref{BMxv}) by Eq. (\ref{BMx}), then the drift term $F(x)/m\gamma$ must be approximately constant over the distance that the velocity is damped. One finds that the associated `length scale' $L$ over which the velocity is damped is simply the pre-factor of diffusion term in Eq. (\ref{BMx}), i.e., $L=\frac{k_b T}{m \gamma}$ \cite{hanggi1995colored}. Enforcing the requirement that $F(x)$ is slowly varying over this length scale we find the inequality $L|F'(x)|\ll|F(x)|$, explicitly:
\begin{align}
    \frac{m \gamma}{k_b T}\gg \left|\frac{F'(x)}{F(x)} \right|,
\end{align}
which must be satisfied for our one variable FPE to be a good approximation.

We now return to our colored noise problem and recall the set of SDEs that define our system:
\begin{align}
    \frac{dn}{dt} &= h(n) + F(n)\eta + g_2(n)\Gamma(t),\\
    \frac{d\eta}{dt} &= -\frac{1}{\tau}\eta + \frac{1}{\tau}\theta(t),
\end{align}
where all functions of $n$ and $t$ are defined in Section \ref{sec:cnpp}. This set of SDEs has an equivalent bi-variate (Stratonovich) FPE given by:
\begin{align}\label{multiFPE}
    \frac{\partial P(n,\eta;t)}{\partial t} = -\frac{\partial}{\partial n}\Big(\big(h(n) +F(n)\eta-\frac{1}{2}g_2(n) & g'_2(n) \big)P \Big)+\frac{1}{\tau}\frac{\partial}{\partial \eta}(\eta P)\\\nonumber &+\frac{1}{2}\frac{\partial^2}{\partial n^2}\left(g_2(n)^2 P\right)+\frac{1}{\tau}\frac{\partial^2}{\partial \eta^2}P.
\end{align}
We now recall Eq. (\ref{eq:prodFP}), i.e., our UCNA approximated one variable FPE, where $\eta$ was adiabatically eliminated:
\begin{align}\label{onevarFPE}
    \frac{\partial P(n;t)}{\partial t} = -\frac{\partial}{\partial n}\left[\left(\Tilde{h}(n)+\Tilde{g}(n)\Tilde{g}'(n)\right)P(n,t)\right] + \frac{\partial^2}{\partial n^2}\left[\Tilde{g}(n)^2P(n,t) \right].
\end{align}
Analogously to the case of the Brownian particle, this one variable FPE can only be approximately correct where the variation of the drift term with respect to the length scale of colored noise fluctuations is small. From Eq. (\ref{eq:eppPreLE}), we identify our length scale as the pre-factor of the noise term whose origin is the adiabatic elimination of $\eta$, i.e.,
\begin{align}
    L=\frac{F(n)}{C(n,\tau)}.
\end{align}
Hence, $C(n,\tau)$ must satisfy the following length scale condition
\begin{align}\label{cond2again}
    C(n,\tau)\gg F(n)\left|\frac{\partial_n \left(\Tilde{h}(n)+\Tilde{g}(n)\Tilde{g}'(n)\right)}{\Tilde{h}(n)+\Tilde{g}(n)\Tilde{g}'(n)}\right|,
\end{align}
if Eq. (\ref{onevarFPE}) is to be a good approximation of Eq. (\ref{multiFPE}), as seen in Eq. (\ref{cond2eq}) from the main text. Note that one can also make the argument that the diffusion term $\Tilde{g}(n)^2$ should also slowly vary with respect to $L$. However, we generally find that this is satisfied if Eq. (\ref{cond2again}) is satisfied, and hence we do not include this as an additional condition on the validity of the UCNA.
\bibliographystyle{unsrt}
\bibliography{biblio.bib}
\end{document}